\documentclass{article}
\usepackage[utf8]{inputenc}
\usepackage{centernot}
\usepackage{arxiv}
\usepackage{bm}
\usepackage[T1]{fontenc}    
\usepackage{url}            
\usepackage{booktabs}       
\usepackage{amsfonts}       
\usepackage{nicefrac}       
\usepackage{microtype}      
\usepackage{lipsum}
\usepackage{graphicx}
\usepackage{verbatim}
\usepackage{latexsym}
\usepackage{subcaption}
\usepackage{setspace}
\usepackage{float}
\usepackage{amsmath}
\usepackage{braket}
\usepackage{mathtools}
\usepackage{tensor}
\usepackage[]{algorithm2e}
\usepackage{xcolor}
\usepackage[normalem]{ulem}
\usepackage{wrapfig,lipsum}
\usepackage{multicol}
\title{Structural importance and evolution: an application to financial transaction networks}
\usepackage{authblk}

\author[1,2, *]{Isobel Seabrook}
\author[1]{Paolo Barucca}
\author[1, 3, 4]{Fabio Caccioli}
\affil[1]{Department of Computer Science, University College London, London, United Kingdom}
\affil[2]{Financial Conduct Authority, London, United Kingdom}
\affil[3]{Systemic Risk Centre, London School of Economics, London, United Kingdom}
\affil[4]{London Mathematical Laboratory, London, UK}
\affil[*]{Correspoding author - ucabeas@ucl.ac.uk}
\date{November 2020}

\begin{document}

\maketitle
\begin{abstract}
A fundamental problem in the study of networks is the identification of important nodes.
This is typically achieved using centrality metrics, which rank nodes in terms of their position in the network. This approach works well for static networks, that do not change over time, but does not consider the dynamics of the network. 
Here we propose instead to measure the importance of a node based on how much a change to its strength will impact the global structure of the network, which we measure in terms of the spectrum of its adjacency matrix. 
We apply our method to the identification of important nodes in equity transaction networks and show that, while it can still be computed from a static network, our measure is a good predictor of nodes subsequently transacting. This implies that static representations of temporal networks can contain information about their dynamics.

\end{abstract}

\section*{Highlights}
\begin{itemize}
    \item We present a measure of node importance that takes into account  overall network structure.
    \item Provides good indication of importance, outperforms other
    measures in predicting subsequent transactions.
    \item Findings provide a novel insights into transaction network behaviour.
\end{itemize}
\twocolumn
\section{Motivation}
Regulating and developing policies for financial markets require the ability to identify important players and to understand how their actions affect other market participants as well as the evolution of the system as a whole. 
This can be achieved by representing interactions between market participants in the form of networks of the transactions they undertake \cite{bardoscia_2021,Schweitzer422}. 
Important nodes can then be identified by considering concepts such as  `centrality', for which there are a number of measures to rank nodes according to their position in the network \cite{Wang, Helander}. 

Markets are often characterised by a wide range of different participant behaviours, manifesting in transaction networks displaying complex structures with a small number of nodes acting as hubs, disconnected communities, and wide ranges of transaction values and trading frequencies \cite{FANG201839, khashanah_2016}. 
For a measure of node importance to provide useful insight to policy makers, it needs to account for these complexities. Furthermore, we need a measure that provides information on how the network would react to changes in the node's activity. 
For this reason, in this study we derive a measure that can be calculated from a static snapshot of a temporal network, but which considers how a change in a node's strength would affect the subsequent structure of the network, which we characterise in terms of the eigenvectors and eigenvalues of the network's weighted adjacency matrix. 
We show that our measure of importance can be used for networks with complex community structures and high heterogeneity of nodes' strengths, and that the measure provides an indication of importance in financial transaction networks, where a key concern is the impact an individual would have on the system if it becomes unable to continue its current level of activity. 

To bring us a step closer to understanding real networks and their stability, we also pose the question of whether important nodes are more or less likely to continue their market participation in a dynamical setting. 
We address this through the application of our methods to networks formed from the transactions of individual stocks (equities) traded on the UK capital markets. We consider daily snapshots of transaction networks and perform a classification exercise 
using a logistic regression model to predict which nodes will appear in the next snapshot given the historical behaviour of the network. This model is a probabilistic model which describes the probability of a node to transact at the next timestamp, so is contributing to the growing body of research exploring the temporal aspects of financial networks.
Our results show that the measure of node importance we propose in this paper can predict nodes being present in the next time snapshot better than other importance measures, including two widely used measures of centrality and the frequency of a node's previous transactions. 

These results indicate that in the context of these equity networks, defining `importance' in terms of how a change occurring will affect the subsequent network structure whilst accounting for communities and disconnected components provides useful insights into the role of network structure in the evolution of these networks. 
This highlights the importance of further research in this area to further understand how network structure relates to stability, particularly in the context of financial networks.

\section{Literature Review}
We aim to provide an approach to node importance which is able to capture the potential of a node to affect the network structure, and to provide insights into the expected behaviour of important nodes. To do this, we need to ensure that our measure can account for complex community structures. We thus explore the links between community detection and node centrality, with a particular focus on methods involving the use of network spectra.

 Many algorithms for community detection have been developed over the past two decades and several comprehensive reviews of the existing methods exist (see for example \cite{JAVED201887, Yang_2016}). However, no single method is found to outperform on all types of networks, with different algorithms presenting different pros and cons depending on the characteristics of the networks being considered \cite{plantie2013survey}. Of relevance for the development of our methods is that several methods of community detection rely on the identification of nodes or edges with high centrality, for example the Girvan-Newman algorithm, which follows an iterative approach to removal of edges with a high betweenness centrality \cite{Girvan7821}, and also Fortunato et. al., who take a similar approach to remove edges with the highest information centrality \cite{fortunato_2004}. In the financial literature, Chan-Lau et. al. \cite{chanlau_2018} explore both community and centrality methods to identify nodes that are systemically `too interconnected to fail' or `too important to fail', and they note that the two complement each other for assessing systemic risk in financial networks. It is also worth noting how both centrality and community detection can be intuitively considered using the concept of a random walk, with a number of methods for community detection, and also for finding central nodes, being defined from the perspective of a random walk and correspondingly defined in terms of the network eigenspectrum. For example, spectral partitioning,  a widely used method for finding communities in graphs, can be interpreted as trying to find a partition of the graph such that a random walk will stay long within the same cluster and rarely move between clusters \cite{Luxburg_2007}. Eigenvector centrality also relates to a random walk of infinite length, in which each node is chosen uniformly at random from the set of neighbours of the current node. Specifically, the eigenvector centrality of a node is proportional to the frequency with which a node is visited during such a walk \cite{Estrada_2010}. When using eigenvector centrality in networks with complex structures, care must be taken to account for disconnected communities, as the measure makes use of the leading eigenpair only. 
 When multiple disconnected components are present in a network, the adjacency matrix can be written in block diagonal form with the eigenspectrum decomposing into the spectra of the individual blocks. This means that the leading eigenpair of the full network will be the leading eigenpair of the largest block, meaning that nodes in a smaller disconnected community will have an eigenvector centrality of 0, even if they play a central role within their community. Katz centrality is a widely used method which accounts for this by adding a free centrality to each node\cite{Katz1953}. Other methods such as those presented in Anguzu et. al. \cite{anguzua2021eigenvector} simply calculate the eigenvector centralities of the network components separately and weight these appropriately. In this paper, we consider whether, in addition to the leading eigenvector, we can make use of the others to account for community structure.

  To establish if we can use the eigenspectrum to account for community structure when determining the importance of a node, we now explore how other researchers have used different parts of eigenspectra to understand network structure.
  
 Much of the research in this area has focused on spectral partitioning methods which make use of the eigenvector corresponding to the second smallest eigenvalue of the Laplacian, also known as the Fiedler vector, to partition graphs \cite{Fiedler1973}. These methods make use of the difference between the coordinates of the Fiedler vector, which provide information about the distance between nodes \cite{Vol_1998}.  However, as is noted in Newman et. al. \cite{Newman_2006}, these methods are still limited to just one part of the spectrum and fail in the detection of community structure when many communities are present. Newman instead gives methods for detecting communities and presents a new idea of `community centrality', by making the observation that modularity can be expressed in terms of the eigenvalues and eigenvectors of the modularity matrix. They take an approach similar to that used in spectral partitioning to maximise the modularity benefit function and show that the eigenvalues of the modularity matrix relate to the community structure. They further show that negative eigenvalues can be used to indicate bipartivity, as well as presenting methods to evaluate network correlations, such as assortativity, using the modularity matrix. 

Given our aim to develop methods that help us understand node behaviour in a dynamical setting, it is also relevant to explore literature that links network eigenspectra to the dynamics of networks. From the perspective of networks' community structure, the concept of `dynamical influence' is explored from the angle of the network's eigenspectrum by Clark et. al. \cite{clark_2019}, who present methods to find `Communities of Dynamical Influence' by investigating the relationships between a system's most dominant eigenvectors. The concept of `temporal centrality' has also been a recent focus of many studies, the majority of which focus on defining temporal random walks to generalise static measures of centrality that are based around the concept of a random walk \cite{ALSAYED201535, estrada_2013, grindrod_2014, grindrod_2011, pan_2011, Rocha_2014}, to produce measures which respect the ordering of events in temporal networks and take into account the temporal distance between events. These methods have recently begun to be applied in a temporal context, such as Zhao et. al. \cite{ZHAO20181104}, who make use of temporal centrality to select peripheral stocks to construct risk diversified portfolios with high return and low risk. Taylor et. al. \cite{taylor_2015} approach things a little differently, presenting a method to extend eigenvector centrality-based methods to temporal networks by coupling centrality matrices for different temporal layers into a supra-centrality matrix. This allows the authors to calculate both the joint centrality for node $i$ at time $t$, as well as marginal and conditional centralities, which allows the study of the node (or temporal layer) centralities separately and the analysis of the centrality trajectory across time. In a similar vein to our research, Kim et. al. \cite{KIM2012983} focus on centrality prediction in dynamic networks, first finding that node centrality is predictable in the context of human social behaviour, before presenting several prediction functions that are suited for different applications. Our findings that node importance is predictive of future presence complements their findings. By studying how static measures of importance relate to future activity, our work is a step towards connecting the static properties to the dynamics of the network. 

Taking into account these examples of the entire network spectrum and its relevance to network community structure and dynamics, we first provide a measure of importance for nodes based on the spectrum. Then we look at whether this measure is predictive of nodes being present in the subsequent snapshot in the context of equity transaction networks. 
This allows us to understand whether we would expect important nodes in these networks to show lower or higher activity. 
This in turn will help us to understand the roles that nodes of differing importance play in establishing the stability of these systems as a whole. 

A key thing to highlight is the simplicity of our methods - both in their use of the spectrum of the adjacency matrix itself, and in the use of snapshots to capture temporal information. Moreover, the results we now present are significant and meaningful despite this simplicity, suggesting that we have uncovered fundamental findings about the behaviour and evolution of financial networks.

\section{Proposed method}
\subsection{Defining structural node importance}

As shown in \cite{seabrook2020}, we can make use of the derivative of a network's leading eigenvalue with respect to adjacency matrix components as a measure of edge importance:
\begin{equation}
   l_e = \frac{\partial\lambda}{\partial A_{ij}}=2x_{0,i}x_{0,j}
    \label{undirected intermediate}
\end{equation}

where $e \equiv ij$ denotes each edge, $\lambda$ refers to the leading eigenvalue, $A_{ij}$ is $ij$th component of the (weighted) adjacency matrix, and $x_{0,i}$ is the $i$th component of the eigenvector corresponding to the leading eigenvalue. This was derived considering small perturbations to the adjacency matrix. Through application of the chain rule, we can derive measures for structural node importance that approximate the derivative of the eigenvalue of the adjacency matrix with respect to an individual node's strength, where a node's strength $S_i$ is the sum of the weights attached to that node\footnote{In an unweighted network, node strength is equivalent to node degree.}. We do this below for undirected and directed networks respectively\footnote{In the applications section, we consider the undirected case only.}. 

\subsubsection{Undirected case}

To derive the equivalent to equation \ref{undirected intermediate} for node importance, we can again consider perturbations to the adjacency matrix to find the derivative with respect to node strength, $\frac{\partial\lambda}{\partial S_i}$. However, in contrast to \cite{seabrook2020}, our perturbation now consists of adding a fixed amount to each node's strength:
\begin{equation}
    S_i \rightarrow S_i + \epsilon
\end{equation}
For this to occur, the change to a node's strength is distributed across its edges. So now, if we consider the perturbation to the adjacency matrix, 
\begin{equation}
    A_{ij} \rightarrow A_{ij}+\epsilon V_{ij}
\end{equation}
where 
\begin{equation}
    V_{ij}=
    \begin{cases}
    \frac{A_{ij}}{S_k} & \textrm{ if } i=k \textrm{ or } j=k\\ 
    0 & \textrm{ otherwise}
    \end{cases}
    \label{pert}
\end{equation}

The perturbation approach then proceeds as follows: First, consider a perturbation to the adjacency matrix $\mathbf{A}$:
\begin{equation}
    \mathbf{A} \rightarrow \mathbf{A}+\epsilon\mathbf{V}
\end{equation}
and the resulting first order changes to the leading eigenvalue $\lambda$ and the associated eigenvector $\ket {x}$ \footnote{Note that we have switched to Dirac notation for conciceness for the rest of the derivation.}:
\begin{equation}
    \lambda = \lambda_0 +\epsilon\lambda
\end{equation}
\begin{equation}
    \ket {x} = \ket {x}_0+\epsilon\ket {x}_1
\end{equation}
Substituting these into our eigenvalue equation
\begin{equation}
\begin{aligned}
(\mathbf{A}+\epsilon\mathbf{V})(\ket {x}_0+\epsilon\ket {x}_1)\\
={}(\lambda_0+\epsilon\lambda_1+...)(\ket {x}_0+\epsilon\ket {x}_1+...)
\end{aligned}
\end{equation}
and considering terms up to 1st order in $\epsilon$
\begin{equation}
\begin{aligned}
    \mathbf{A}\ket {x}_0+\epsilon\mathbf{V}\ket {x}_0+\epsilon\mathbf{A}\ket {x}_1\\
    =\lambda_0\ket {x}_0+\epsilon\lambda_1\ket {x}_0+\epsilon\lambda_0\ket {x}_1
    \label{lin_approx}
\end{aligned}
\end{equation}
Then we can consider each of the terms in $\epsilon^n$ separately,
\begin{equation}
    \epsilon_0: \textbf{A}\ket {x}_0=\lambda_0\ket {x}_0
\end{equation}
\begin{equation}
    \epsilon_1: \textbf{V}\ket {x}_0+\mathbf{A}\ket {x}_1=\lambda_1 \ket {x}_0+ \lambda_0\ket {x}_1
\end{equation}
By multiplying the equation for $\epsilon^1$ by the left eigenvector $\prescript{}{0}{\bra{x}}$ and making use of the hermitian properties of $\mathbf{A}$ such that $\prescript{}{0}{\bra{x}}\mathbf{A}=\lambda_0\prescript{}{0}{\bra {x}}$, we find
\begin{equation}
    \tensor*[_0]{\braket{x|\mathbf{V}|x}}{_0} = \lambda_1\tensor*[_0]{\braket{x|x}}{_0} 
\end{equation}

If we expand the indices of this and consider our specific perturbation in equation \ref{pert}, 
\begin{equation}
\begin{aligned}
    \sum_{ij} x_{0,i} V_{ij}x_{0,j}\\ =\sum_{ij}x_{0,i} \frac{A_{ij}}{S_k}x_{0,j} \delta_{ik}+ \sum_{ij}x_{0,i} \frac{A_{ij}}{S_k}x_{0,j} \delta_{kj} \\ = \frac{2}{S_k} \sum_{j}x_{0,k} A_{kj}x_{0,j}
    \end{aligned}
 \end{equation}
where we have evaluated the delta terms and relabelled the indicies. From, this, we can find the derivative of the eigenvalue with respect to node strength: 
\begin{equation}
    \frac{\partial \lambda}{\partial S_i} = \frac{\partial\lambda}{\partial(\sum_j A_{ij})} = \frac{2}{S_i}\sum_j x_{0,i}A_{ij}x_{0,j} 
\end{equation}

\subsubsection{Directed case}
In the case of a directed network, $\mathbf{A}$, the perturbations to the matrix either correspond to changes to in strength or out strength and we do not need to perturb the matrix symmetrically. Further to this, in contrast to the above, $\mathbf{A}$ is not hermitian and so we cannot use that $\mathbf{x}^T\mathbf{A}=\lambda \mathbf{x}^T$. However, the matrix product $\mathbf{M}=\mathbf{AA^T}$ is symmetric and hermitian.

Our edge level result for the directed case from \cite{seabrook2020} is 
\begin{equation}
  \frac{\partial s^A}{\partial M_{ij}}=\frac{x^M_{0,i} x^M_{0,j}}{2s^A}
    \label{directed intermediate appx}
\end{equation}

where $x_{0,i}^M$ refers to the ith component of the eigenvector of $\mathbf{M}$ corresponding to the leading eigenvalue of $\mathbf{M}$, which is also also known as the singular value of $\mathbf{A}$, $s^A$. We can again relate to the strength by considering a Taylor expansion of the matrix $\mathbf{A}$
\begin{equation}
    A_{ij}=A_{ij}^0+\epsilon A_{ij}^1+\epsilon^2 A_{ij}^2
\end{equation}
which means that to 1st order, 
\begin{equation}
    M_{ij}=\sum_k (A_{ik}^0+\epsilon A_{ik}^0)(A_{jk}^0+\epsilon A_{jk})
\end{equation}
\begin{equation}
    =M_{ij}^0+2\epsilon M_{ij}^0 
\end{equation}
so
\begin{equation}
    \frac{\partial M_{ij}}{\partial \epsilon}=2M_{ij}
\end{equation}
Which gives our result when applying the chain rule as above
\begin{equation}
    \frac{\partial s^A}{\partial S_i}=\frac{1}{S_i s^A}\sum_j x_{0,k}x_{0,j} M_{ij}
    \label{m_i directed appx}      
\end{equation}
To summarise our final results of these derivations, we present equations \ref{m_i symm main} and \ref{m_i directed main} for undirected and directed networks respectively:

\begin{equation}
    m_i = \frac{\partial\lambda}{\partial S_i} \equiv \frac{2}{S_i}\sum_j x_{0,i} x_{0,j}
    \label{m_i symm main}
\end{equation}
\begin{equation}
    m_i = \frac{\partial s^A}{\partial S_i} \equiv \frac{1}{S_i s^A}\sum_j x_{0,i}x_{0,j} M_{ij}
    \label{m_i directed main}      
\end{equation}
here $m_i$ denotes the importance of node $i$, $S_i$ is the strength of node $i$, $s^A$ is the singular value of the adjacency matrix $A_{ij}$, and $M_{ij}$ is the matrix product $\mathbf{M}=\mathbf{AA^T}$. We note here that our measure of node importance is, by design, inversely proportional to node strength.  Although this is in contrast to measures of centrality, here we are defining importance by considering an individual node experiencing a fixed size change to its strength, meaning that a more connected node will distribute its change across more edges, having a smaller effect on each of its neighbours individually. An alternative definition of importance could consider fixed changes to each edge, effectively producing the inverse of our defined measure. However, for our application to financial transaction networks, it is important to understand the scenario in which a participant in the market experiences a decrease in its available inventory and how this impact will propagate to its neighbours. A well connected node in a network will have the option of spreading this impact across multiple trading relationships, whereas a poorly connected node will present a larger risk to its counterparties. For this reason, in this paper, we define importance from the perspective of fixed changes to node strength. In both the directed and undirected case, it is also worth noting that the derivations can be generalised to allow new links to be added/removed however new nodes cannot be added or removed.

\subsubsection{Extension of node importance method}

\label{sec:comm_struct}
Although in \cite{seabrook2020} we considered only perturbations to the leading eigenvalue and its associated eigenvector, equations \ref{m_i symm main} and \ref{m_i directed main} are relevant for any of the single components of the eigenspectrum.  Later in section \ref{applications} we demonstrate how different parts of the networks considered in this paper relate to different parts of the eigenspectrum and we propose that our methods can be made `structurally aware' through the use of multiple components of the eigenspectrum. First, we note that care must be taken in identifying the relevant eigenvector from the eigenspectrum of the graph. 
\subsection*{\textit{\small Toy network exploration of network spectrum}}
\label{eig_loc}
\begin{figure}
    \centering
    \includegraphics[width=\linewidth]{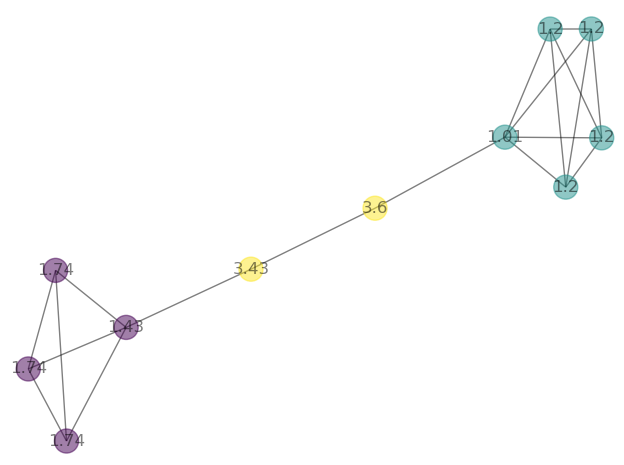}
    \caption{Barbell graph, nodes coloured by result of k-means run on the eigenvectors corresponding to the positive eigenvalues of the adjacency matrix. Nodes are labelled by the value of the measure $m_b$.}
    \label{fig:barbell}
\end{figure}

Here we briefly explore whether the use of multiple components of the network spectra can be used to capture different structures in networks through the use of a toy network. We consider a barbell network with two unevenly sized cliques joined by a bridge, shown in figure \ref{fig:barbell},  in order to observe how the different components of the eigenspectra are relevant for the different communities present in this network.
Table \ref{tab:eigenvectors} shows the eigenvector values corresponding to the 3 positive eigenvalues of the adjacency matrix. If we consider the nodes in the largest clique (top right in figure \ref{fig:barbell}, nodes 6 to 10), we see that the largest eigenvector components are seen for the eigenvector corresponding to the leading eigenvalue (eigenvector 1). Considering nodes 0 to 3 (in the bottom left clique), we see that the largest magnitude eigenvector components are seen for eigenvector 2. The nodes in the bar (nodes 4 and 5) both show the largest component for eigenvalue 3. We further support these observations through the use of a k-means clustering, applied to the three positive eigenvectors, which resulted in the clustering of the nodes shown by the different colours in figure \ref{fig:barbell}, demonstrating that the different eigenvectors have relevance for the different communities present in the network. 

\begin{table}
     \centering
     \begin{tabular}{|c|c|c|c|c|c|}
     \hline
       Node & Eigenvector 1 & Eigenvector 2 & Eigenvector 3  \\
       \hline
        0 &	0.006 & -0.478 & -0.159	 \\
        1  & 0.006 & -0.478 & -0.159\\
        2  & 0.006 & -0.478 & -0.159	 \\
        3  & 0.013 & -0.524 & 0.121	 \\
        4 &	0.033 &	-0.189 & 0.629	 \\
        5  & 0.122 & -0.060 & 0.658	 \\
        6  & 0.463 & 0.002 & 0.187\\
        7  & 0.439 & 0.016 & -0.106 \\
        8  & 0.439 & 0.016 & -0.106\\
        9 & 0.439 &	0.016 &	-0.106 \\
        10 & 0.439 & 0.016 & -0.106 \\
         \hline
     \end{tabular}
     \caption{Eigenvector values corresponding to the three positive eigenvalues for the barbell network displayed in figure \ref{fig:barbell}. Nodes 0 to 3 are the nodes within the top left clique, nodes 4 and 5 make up the bridge and nodes 6 to 10 are the nodes within the bottom right clique.}
     \label{tab:eigenvectors}
 \end{table}
 
Here we have demonstrated that the n'th largest community is found to correspond to the n'th largest eigenvalue and its eigenvector and that the magnitude of the components of this eigenvector for the given community will be larger than the components for the other eigenvectors. So we expect that by taking the largest magnitude eigenvector components corresponding to the nodes in the community as the `correct' eigenvector components to represent the nodes in that community, our measure will be `community aware'. To assess this, we propose extending our structural importance metric to  make use of the spectrum in one of four ways:
\begin{enumerate}
    \item Only make use of the leading eigenvalue and its associated eigenvector in equation \ref{m_i symm main}. This measure is expected to perform well when there are no communities present. We will refer to this as $m_a$.
    \item Identify, for each node, the eigenvector with the largest magnitude component for that node and the eigenvalue associated with this. Use these to compute equation \ref{m_i symm main} for each node. We will refer to this as $m_b$. 
    \item To understand whether node importance has meaningful contributions from all parts of the spectrum, consider importance as the sum of equation \ref{m_i symm main} for all eigencomponents. We will refer to this as $m_c$.
    \item Consider as for $m_c$, but only make use of the part of the eigenspectrum with positive eigenvalues. We will refer to this as $m_d$.
\end{enumerate}

The nodes in figure \ref{fig:barbell} are also labelled with the individual $m_b$ node importances. We see here that the nodes making up the bridge, which themselves have very few connections, are the most important, and the nodes in the larger clique are the least important. This demonstrates that our measure is performing as expected, as a fixed change to a node's strength when the node is present in the bridge would have a larger impact on the rest of the network than for a node in a clique.

\subsection{Node importance and network evolution}
It is intuitive to explore how importance in a static sense relates to future activity in a network, since an importance measure is only useful in practice if it is able to provide actionable information. Several studies have used node or edge importance to explain structural changes in real network systems \cite{Restrepo2, Kim_2012, takes_2016} and in particular financial systems \cite{Battiston, Barucca, Bardoscia}. Also of note are  Battiston et. al. \cite{Battiston} and Markose et. al. \cite{Markose}, which present evidence of organisations being `too central to fail' and `too interconnected to fail' respectively, demonstrating the importance in financial applications of understanding not only how they are systemically connected nodes are, but also how central they are. 

When considering the use of our measures of importance in understanding real networks and their stability, we can consider whether important nodes are more or less likely to be present in the subsequent snapshot given their current importance. To do this, we make use of logistic regression to predict subsequent node presence from historical feature vectors. The logistic regression we consider can be interpreted as a probabilistic model that gives the probability that a node subsequently transacts given historical properties: 
\begin{equation}
    P(Y=1|\mathbf{X}=\mathbf{x}) = \frac{\exp{(\bm{\alpha}+\bm{\beta}^T \bm{x})}}{1+\exp{(\bm{\alpha} +\bm{\beta}^T \bm{x})}}
\end{equation}
Where $Y$ is our target variable taking a binary value per node, $\mathbf{X}$ is a vector of feature values, and $\bm{\alpha}$ and $\bm{\beta}$ are the regression model coefficients. 
The feature vectors consist of our node importance measures, along with eigenvector and pagerank centralities as benchmark measures, and other node level attributes, degree and community. These features are calculated as averages over all the previous time periods in the data available prior to the snapshot in question, to answer the question `given what I know about the network up to today, what do I know about tomorrow?'. We also include a further feature of the number of times that a node has been present in the network prior to the snapshot as a benchmark to compare our measure to. The target variable for the classifier is a binary variable indicating whether a node that is present in the current snapshot is also present in the next snapshot.
 Since we observe fairly high levels of class imbalance across the datasets considered \footnote{Equity-1 shows 1499 of 2063 present in the subsequent snapshot, Equity-2 shows 724 of 880 present, and Equity-3 shows 1803 of 2237 present}, we apply a random over-sampling strategy to correct this for all three datasets. We make use of 5-fold cross validation \footnote{For this, we split our data into training and test sets, with a 40-40-20 train, validation, test split. We split the data whilst keeping the ordering of time, so that the model is not trained on data from the future.} to select the best classifier and its associated parameter values\footnote{Both logistic regression and random forest classifiers were considered to allow for potentially non-linear relationships. In practice, the logistic regression model consistently performed the best, so the results presented in this paper are for logistic regression classifiers.}. This not only allows us to assess whether our measures of node importance are predictive of subsequent node activity, but it also provides us with the means of comparing the different entries of the feature vector through their feature importances. To do this, we make use of permutation importance, which calculates the increase in the model's prediction error after permuting the feature \cite{fisher2019models}. Since this measure is only able to capture the importance of features in a global sense without accounting for the role of a feature in individual predictions, we also make use of Shapley values, which use concepts from co-operative game theory to explain the additional importance of each variable for each individual observation \cite{GIUDICI2021114104}. More specifically, we make use of the SHapley Additive exPlanations (SHAP) approach, which quantifies the contribution each feature in a Machine Learning model makes to the prediction of individual observations \cite{shap}. When using SHAP to explain the probability of a linear logistic regression model, we note that strong interaction effects would affect the performance since the model is not additive in the probability space. To account for this, we use SHAP to explain the log-odds of the model, as there is a linear relationship between the model's inputs and this output \cite{shap}. When evaluating the overall performance of the classifiers, we make use of precision, recall and the area under the Receiver Operating Characteristic curve (ROC AUC). 
 
 We benchmark the results against a null model consisting of the average over 100 trials in which edges are randomly present with probability equal to the fraction of observed edges. For this benchmark, the Confidence Intervals were determined empirically by discarding the top and bottom 5\% of the precision and recall scores and a `coin toss' approach was taken to calculate the Confidence Intervals for the model precision and recall thresholds such that the probability of observed outcome is higher than some threshold $\alpha$. 
The first step of this approach is to note that precision is the probability of ground truth positive from all positive predictions and recall is the probability of a ground truth positive from all correct predictions. We can then make use of the Binomial distribution probability mass function to estimate the probability of the observed outcome depending on the chance of a `positive flip' (the value of precision or recall):
\begin{equation}
    Pr(k=TP; n,p) = \frac{n!}{TP!(n-TP)!} p^{TP}(1-p)^{(n-TP)}
    \label{prob_model}
\end{equation}
where $k$, the number of successes, is equal to the number of True Positives (TP) in both cases, n is the number of True Positives and False Positives (TP+FP) for the case of precision and the number of True Positives and False Negatives (TP+FN) for the case of recall and $p$ is the probability of success. Taking the cumulative of this and inverting for $p$ allows us to find the Confidence Intervals for the Precision and Recall.

In practice, we can make use of the Central Limit Theorem when we have sufficiently large samples, in that the sum of random variables closely follows a normal distribution, allowing us to make use of the confidence intervals of a normal distribution. When comparing to the null model results in this paper, we also present empirically calculated confidence intervals and average precision and recall for our null model, which randomly predicts 1 or 0 in proportion to the dataset prior. 
The Confidence Intervals for the ROC AUC were calculated using a bootstrap approach, with 1000 iterations of random sampling with replacement from the training dataset.

\section{Applications}
\label{applications}
\subsection{Application to individual equity stocks}
The bulk of the results presented in this paper focus on the identification of important nodes in transaction networks of three different equity stocks traded on the UK capital markets, reported under MIFID II regulations. These datasets were available in their raw transaction form, containing information on price, volume, transaction time and anonymized identities of market participants. The data was aggregated daily, from which we constructed networks with the market participants as the network nodes and edges representing the net value traded between two market participants and covers a 5 month period from July 2018. If aggregated across the entire time period, the first of the networks, which we refer to as Equity-1, contains 232 nodes and 6,961 edges. The second, Equity-2, is smaller but much denser containing 94 nodes and 3,684 edges and the third, Equity-3, contains 263 nodes and 9,094 edges. An exploration of the key properties of these networks can be found in the Supplementary Information of \cite{seabrook2020}. 
As these datasets are not open source, later we also include an application of our methods to an open source dataset of inter-country trades in financial services, created by the OECD and WTO \cite{wto_data}.
\subsubsection{Exploration of network community structure}
First, to build our understanding of how the equity transaction networks evolve, we explore how the community structure varies across time and how the distributions of various node level measures differ for nodes that do appear in subsequent snapshots to nodes that don't. The former is explored since we are proposing a measure of node importance that is able to capture community structure, so we first need to verify that the networks we are considering consistently show a community structure. The latter provides us with an indication of which node level measures we might expect to provide us with information on network evolution and also helps us understand whether different classes of nodes are more, or less likely to subsequently transact.

\begin{figure*}
    \centering
\begin{subfigure}{.33\textwidth}
     \centering
    \includegraphics[width=54mm]{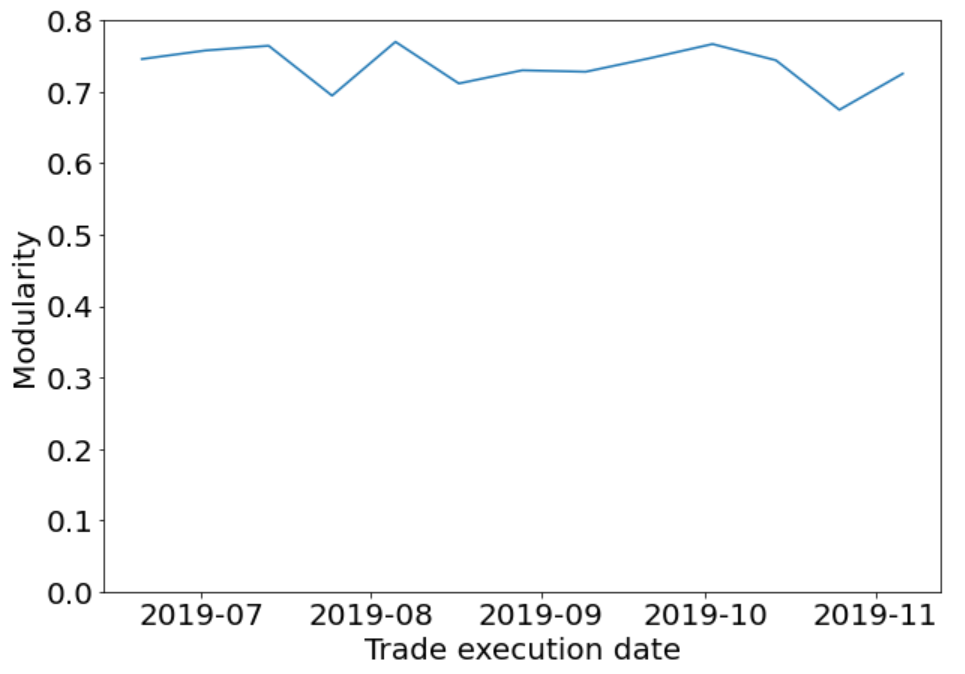}
    \caption{Equity-1}
\end{subfigure}%
\begin{subfigure}{.33\textwidth}    
   \centering
    \includegraphics[width=54mm]{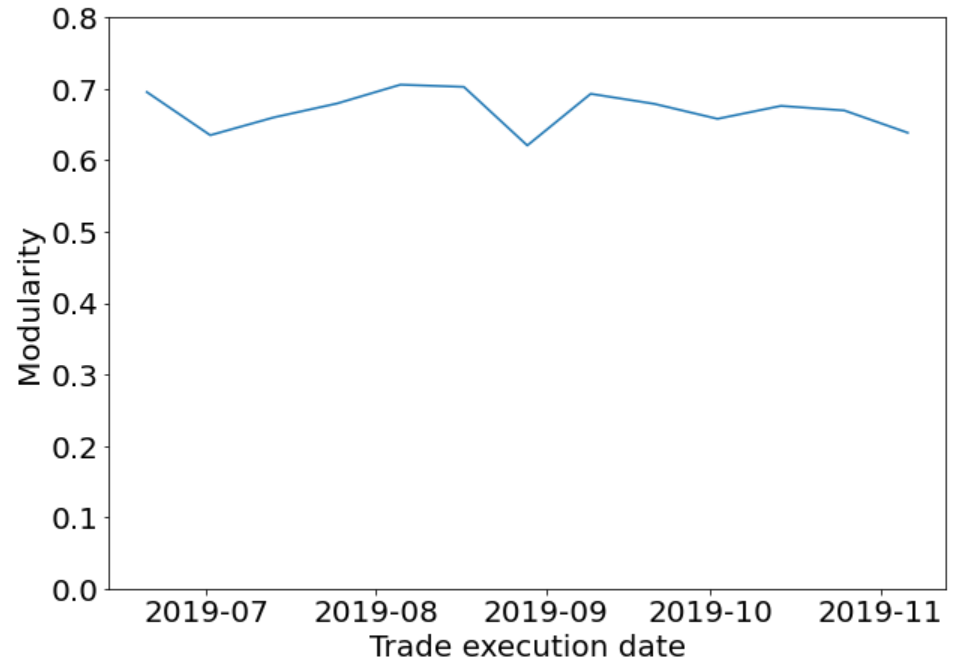}
    \caption{Equity-2}
\end{subfigure}
\begin{subfigure}{.33\textwidth}
   \centering
    \includegraphics[width=54mm]{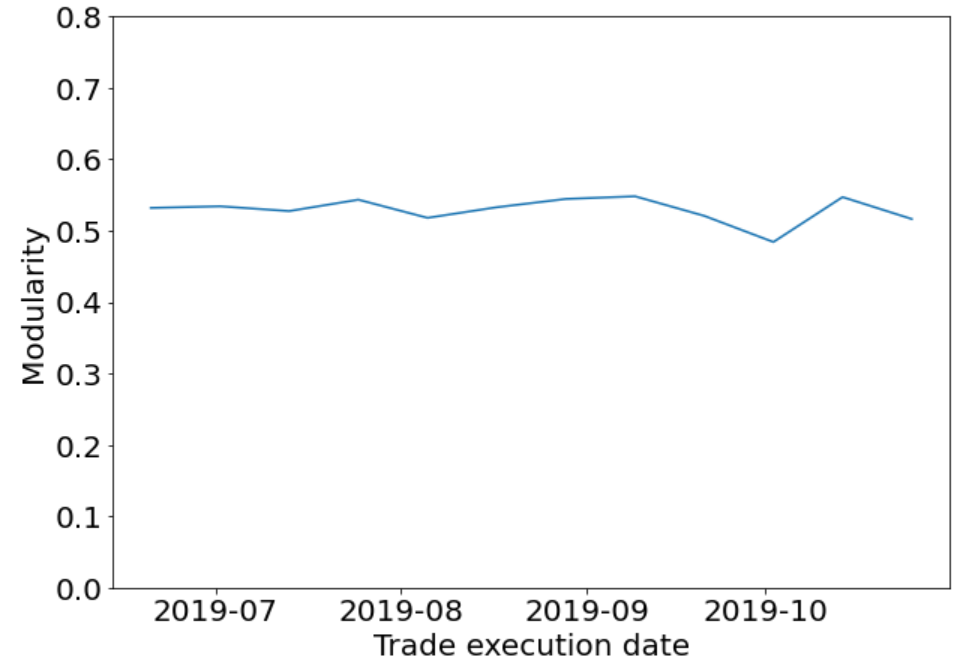}
    \caption{Equity-3}
\end{subfigure}

\caption{Modularity across time for the three different equities networks}
\label{fig:modularity}
\end{figure*}

Figure \ref{fig:modularity} examines how the networks evolve across time, by considering the variation in the modularity \cite{Newman_2004} (the fraction of the edges that fall within the given groups minus the expected fraction if edges were distributed at random). Aggregating on a half-monthly basis to reduce noise corresponding to days of very low trading activity at weekends and public holidays, we see that all three networks have a largely static modularity, with all showing similar average modularity of around 0.4-0.6, suggestive of a meaningful community structure that does not vary significantly across the observation period. The network with the lowest modularity is Equity-3, which is in agreement with what we see in the main text when visually exploring these networks, as this network consists of one large connected component, with the smaller disconnected components observed for the other two networks not present in this dataset.

\begin{figure}
\vspace{-1cm}
    \begin{subfigure}[t]{\linewidth}
        \centering
        \includegraphics[width=0.7\textwidth, height = 0.2 \paperheight]{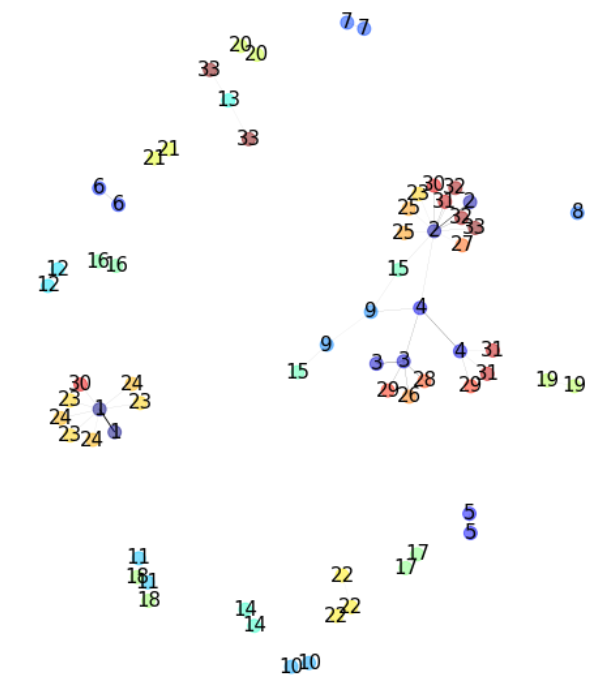}
    \caption{Equity-1}
    \label{fig:eq1_plot}
    \end{subfigure}

    \begin{subfigure}[t]{\linewidth}
        \centering
        \includegraphics[width=0.7\textwidth, height = 0.2 \paperheight]{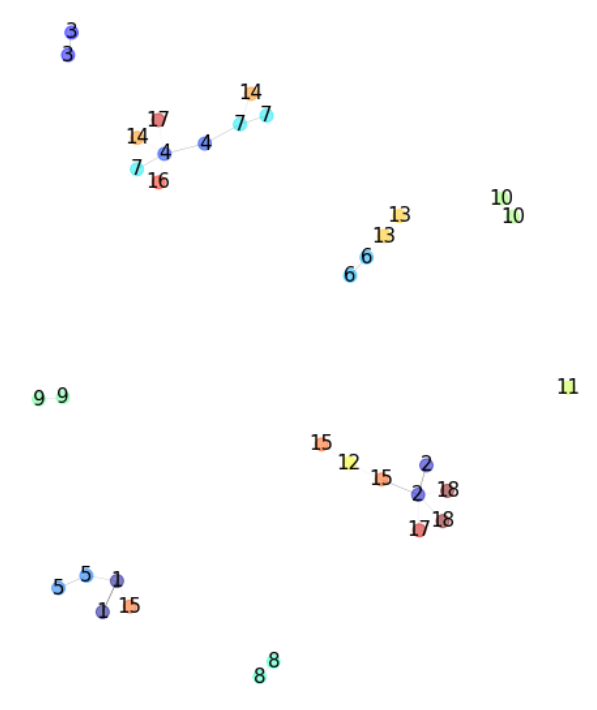}
    \caption{Equity-2}
    \label{fig:eq2_plot}
    \end{subfigure}

    \begin{subfigure}[t]{\linewidth}
        \centering
        \includegraphics[width=0.7\textwidth, height = 0.2 \paperheight]{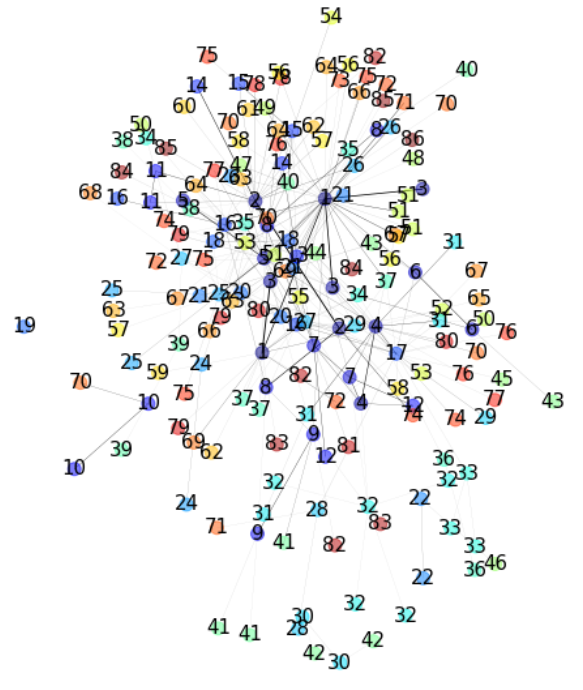}
    \caption{Equity-3}
    \label{fig:eq3_plot}
    \end{subfigure}
    \caption{Initial snapshot networks for the three Equities, colours and numbers representing the ranking of the eigenvalue that corresponds to the eigenvector with the largest magnitude for each node.}
\end{figure}

We can also build our understanding of how the different components of the eigenspectra relate to the communities of the networks by exploring how nodes rank by eigenvalue if we select for each node the eigenvalue with the largest magnitude eigenvector component for that node. In figures \ref{fig:eq1_plot}, \ref{fig:eq2_plot} and \ref{fig:eq3_plot}, the nodes are coloured and numbered by the rank of the eigenvalue that is selected (rank 1 corresponds to the largest eigenvalue). We can see clearly in all networks that nodes within small communities often, but not exclusively, select the same eigenvalue and that nodes playing similar roles within the network show similar ranks for their eigenvalue. Where nodes in the same connected component select different eigenvalues, hub nodes select higher ranked eigenvalues. This suggests that if we make use of our measure of importance in equation \ref{m_i symm main} whilst selecting the most relevant eigencomponent, this would assign a larger importance to these nodes. However, this will be partially counteracted by the inverse strength factor in our structural importance measure, which makes sense since a node in a small but well connected community has few direct neighbours to spread the impact of a change in strength between but will have a high reachability to other nodes overall.

\begin{figure*}
    \centering
\begin{subfigure}{\textwidth}
     \centering
    \includegraphics[width=0.9\linewidth]{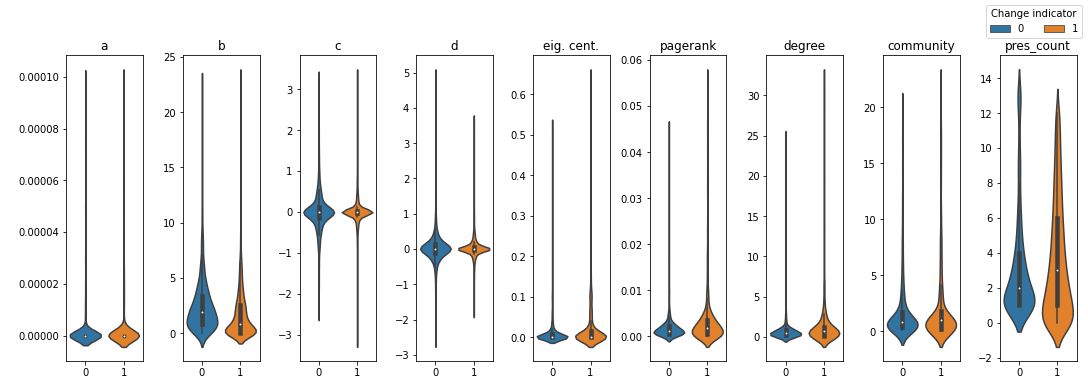}
    \caption{Equity-1}
    \label{fig:box_1}
\end{subfigure}%

\begin{subfigure}{\textwidth}    
   \centering
    \includegraphics[width=0.9\linewidth]{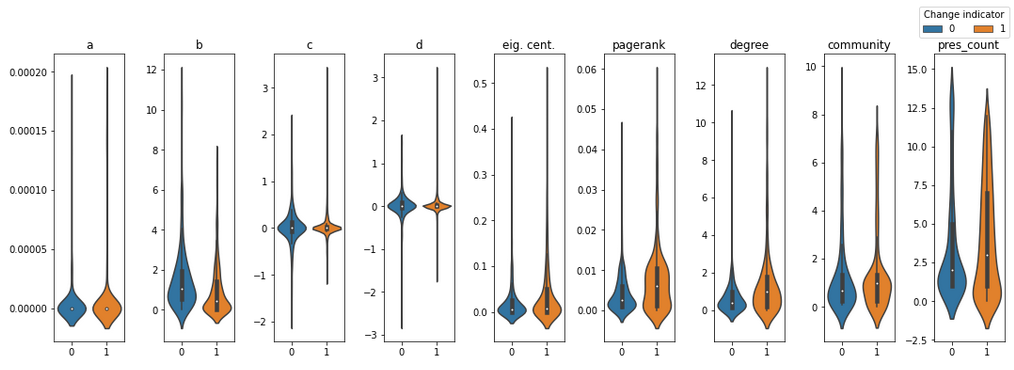}
    \caption{Equity-2}
    \label{fig:box_2}
\end{subfigure}

\begin{subfigure}{\textwidth}
   \centering
    \includegraphics[width=0.9\linewidth]{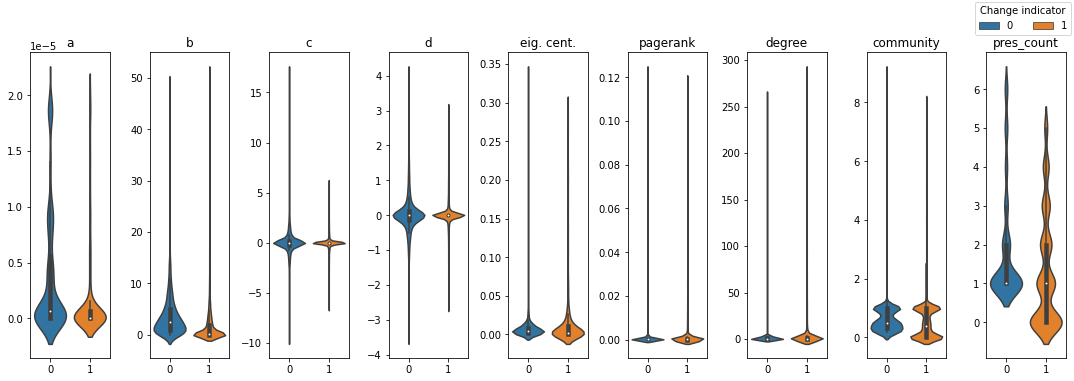}
    \caption{Equity-3}
    \label{fig:box_3}
\end{subfigure}
\caption{Distributions of the different node importance measures across nodes which are subsequently present in comparison to those that aren't.}
\label{fig:boxes}
\end{figure*}

Figure \ref{fig:boxes} shows the distributions of the values of $m_{a-d}$ computed according to the 4 different eigenvalue inclusion schemes along with two benchmark measures of node importance, pagerank and eigenvector centrality and also degree, community label and the number of times a nodes has been present in the historical data (presence count). We use violin plots to present the distributions, which show the kernel density estimated distribution plotted on top of a boxplot showing the mean and interquartile range. The plots are split by whether or not nodes are subsequently present in the network. We see here that $m_b$, which selects the relevant eigenvalue component for each node, visually shows the largest difference in the distribution mean for present nodes in comparison with absent nodes across the three datasets. We also observe that nodes that are subsequently present are observed with smaller values of $m_b$, in contrast to eigenvector centrality and pagerank, which both show changing nodes having slightly larger values. Table \ref{tab:p-values} shows the p-values for a two-sided t-test for the differences in the mean values for change vs. no change for each of the different measures. We see that $m_a$ $m_c$, and $m_d$ do not show $p<0.01$ \footnote{Due to the comparisons of 9 metrics simultaneously, we have applied a Bonferroni correction to the p-value threshold.} for all datasets. This non-significant difference in the mean values for changing vs. unchanging nodes suggests that we would not expect these measures to be predictive of subsequent change. On the other hand, $m_b$, community, degree, eigenvector centrality, pagerank and presence count all show significant differences in the mean values for all datasets, making these measures better candidates for prediction of subsequent node presence.   

\begin{table}
    \centering
    \begin{tabular}{|c|c|c|c|}
    \hline
        Measure & Equity-1 & Equity-2 & Equity-3  \\
        \hline
        $m_a$ & $3.23 \times 10^{-11}$ & $8.17\times 10^{-2}$ & $5.82\times 10^{-63}$ \\
        $m_b$ & $1.34 \times 10^{-4}$ & $3.96\times 10^{-13}$ & $2.42 \times 10^{-135}$ \\
        $m_c$ & $1.18\times 10^{-1}$ & $8.47\times 10^{-1}$ & $3.62 \times 10^{-1}$ \\
        $m_d$ & $5.09 \times 10^{-1}$ & $8.23\times 10^{-1}$ & $4.62\times 10^{-1}$ \\
        Community & $1.76\times 10^{-21}$ & $1.03\times 10^{-22}$ & $2.57\times 10^{-2}$ \\
        Degree & $9.53\times 10^{-49}$ & $2.36\times 10^{-13}$ & $4.62\times 10^{-10}$\\
        Eig. cent. & $4.73\times 10^{-37}$ & $42.99\times 10^{-8}$ & $9.89\times 10^{-12}$ \\
        Pagerank  & $3.51\times 10^{-58}$ & $1.56\times 10^{-14}$ & $5.25\times 10^{-5}$ \\
        Pres. count & $3.72 \times 10^{-53}$ & $1.22 \times 10^{-2}$ & $9.78 \times 10^{-52}$\\
        \hline
    \end{tabular}
    \caption{p-values for a two-sided t-test for the differences in the mean values for nodes which change and nodes which don't change for each of the different node level measures, for the Equity transaction datasets.}
    \label{tab:p-values}
\end{table}

\begin{figure*}
    \centering
\begin{subfigure}{.4\textwidth}
     \centering
    \includegraphics[width=\linewidth]{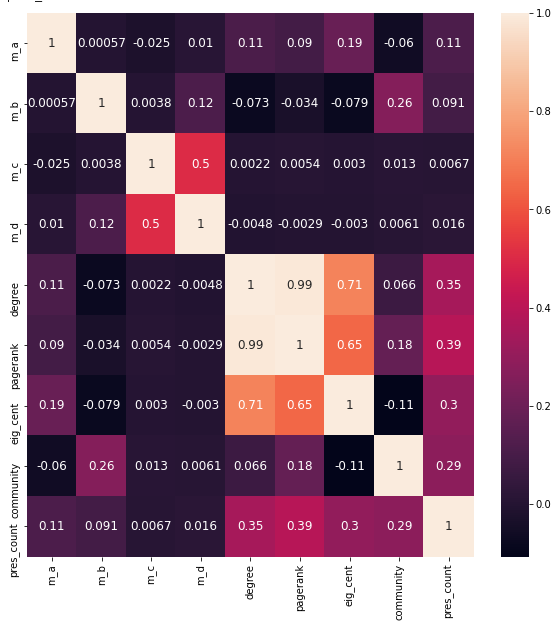}
    \caption{Equity-1}
    \label{fig:eq1_corr}
\end{subfigure}%
\begin{subfigure}{.4\textwidth}    
   \centering
    \includegraphics[width=\linewidth]{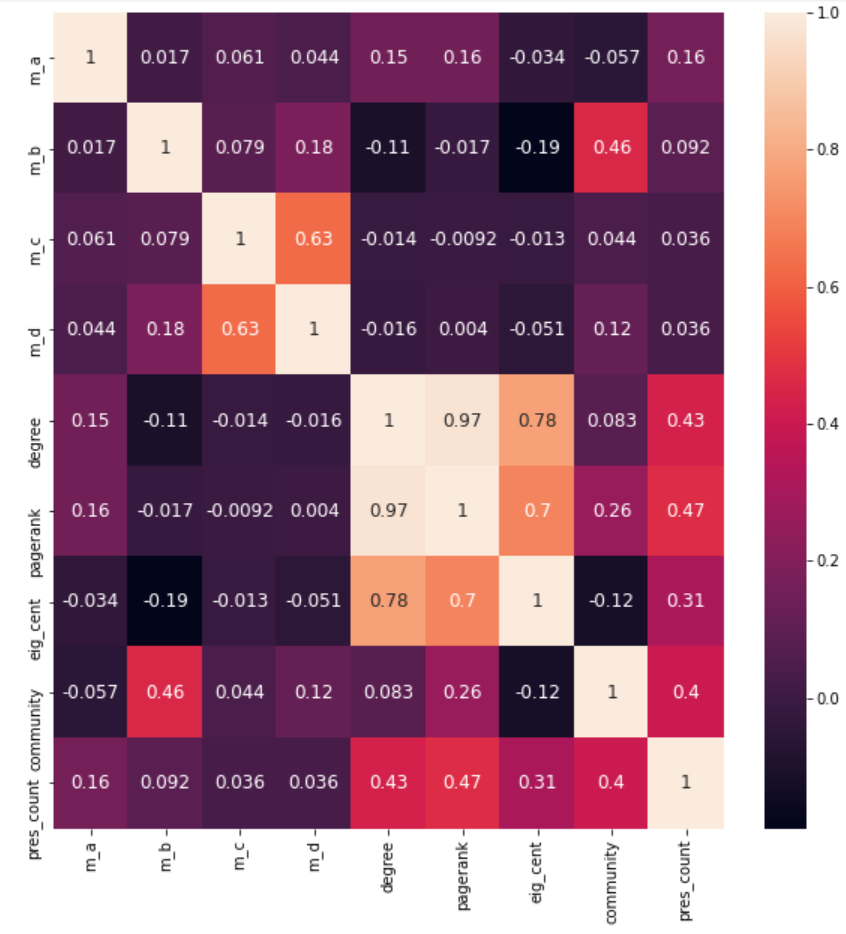}
    \caption{Equity-2}
    \label{fig:eq2_corr}
\end{subfigure}
\begin{subfigure}{.4\textwidth}
   \centering
    \includegraphics[width=\linewidth]{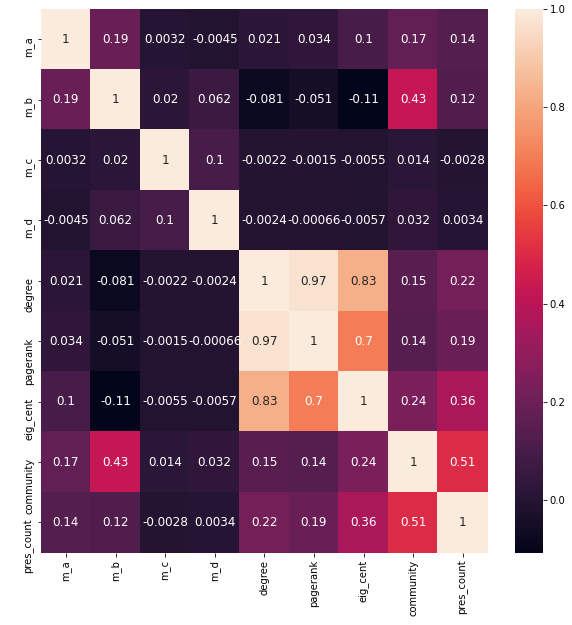}
    \caption{Equity-3}
    \label{fig:eq3_corr}
\end{subfigure}
\caption{Pearson correlations for the different equity transaction datasets}
\label{fig:equity_corrs}
\end{figure*}

In order to assess the similarities between the different measures and to ensure that our model is not impacted by large correlations between the features, we consider the Pearson correlations between the rankings of nodes according to the different measures, shown in figure \ref{fig:equity_corrs}. 
In general across all three datasets, we see that the measures $m_a$, $m_b$, community and presence count show no significant correlations with any other measures. For equity-1 and equity-2, $m_c$ and $m_d$ are moderately correlated with each other, which is expected since the two measures differ only in their use of the part of the spectra with negative eigenvalues for which the eigenvector components will be small. For equity-1, high correlations were observed between degree and both pagerank and eigenvector centrality, so degree was not included in the feature vector for the classifier for this dataset. For the other two datasets, high correlations were observed between pagerank, degree, and eigenvector centrality, so both pagerank and degree were not included in the feature vector for the classifiers for these datasets. These large correlations are indicative of the dominance of hub nodes in these networks.

\subsubsection{Prediction experiments}
\begin{figure*}
    \centering
\begin{subfigure}{.33\textwidth}
     \centering
    \includegraphics[width=\textwidth]{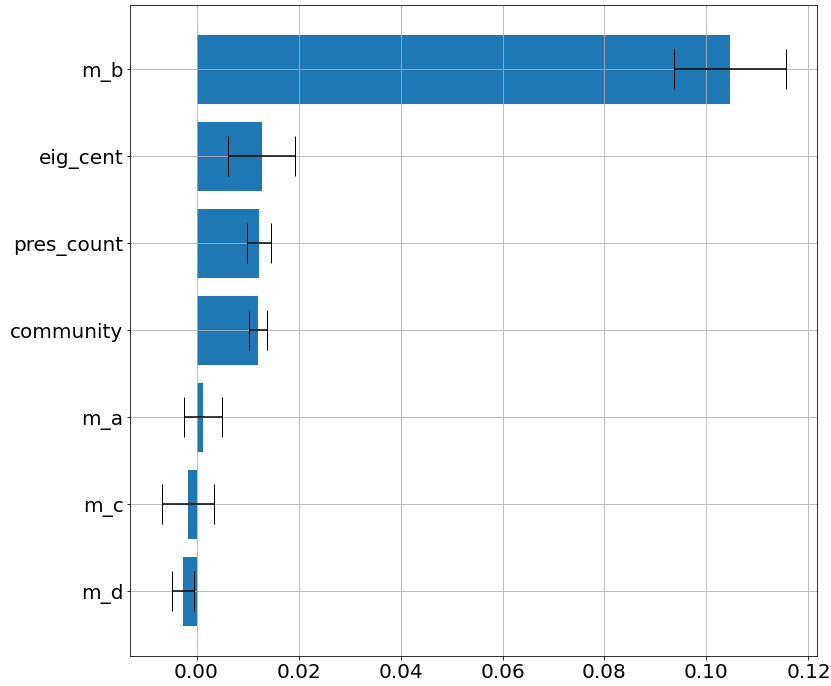}
    \caption{Equity-1}
    \label{fig:fi_1}
\end{subfigure}%
\begin{subfigure}{.33\textwidth}    
   \centering
    \includegraphics[width=\textwidth]{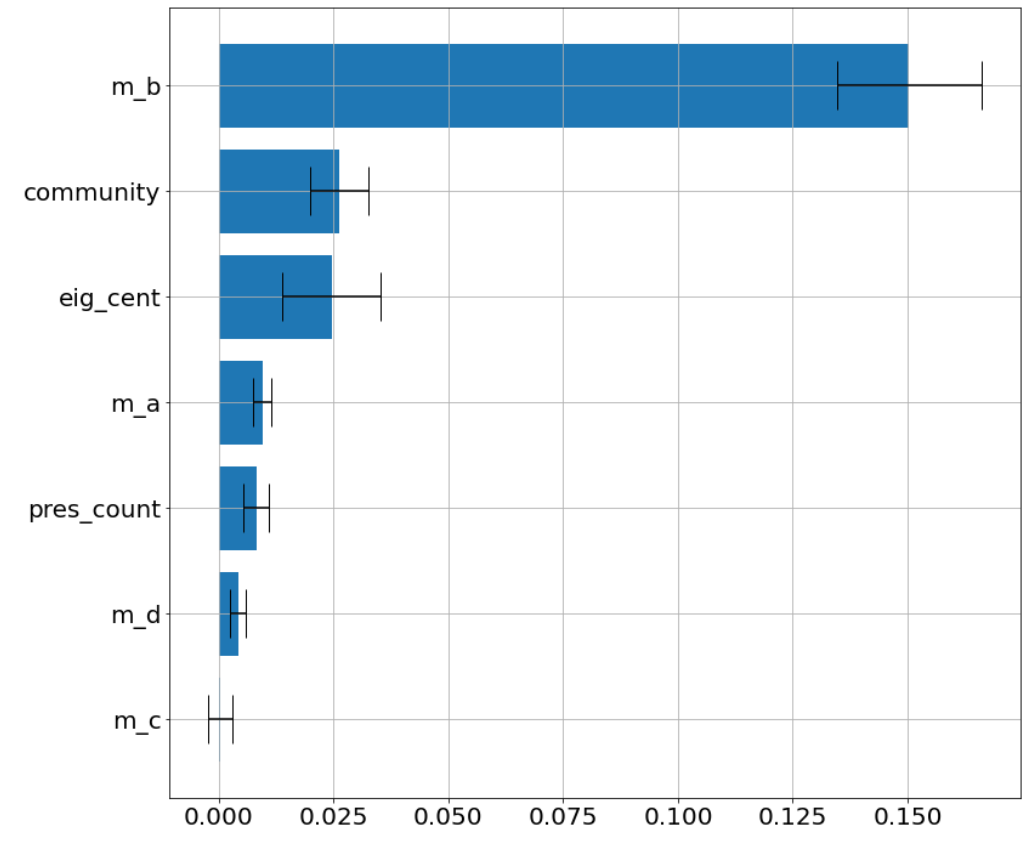}
    \caption{Equity-2}
    \label{fig:fi_2}
\end{subfigure}
\begin{subfigure}{.33\textwidth}
   \centering
    \includegraphics[width=\textwidth]{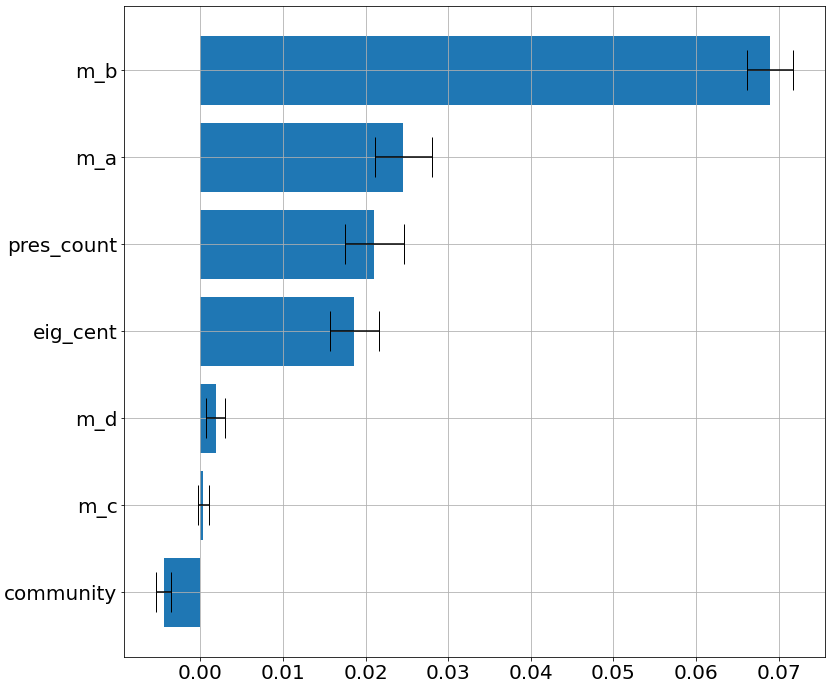}
    \caption{Equity-3}
    \label{fig:fi_3}
\end{subfigure}

\caption{Logistic regression permutation importance for the different node importance measures.}
\label{fig:feature_imps}
\end{figure*}

\begin{figure*}
    \centering
\begin{subfigure}{\textwidth}
    \centering
    \includegraphics[width=0.7\textwidth]{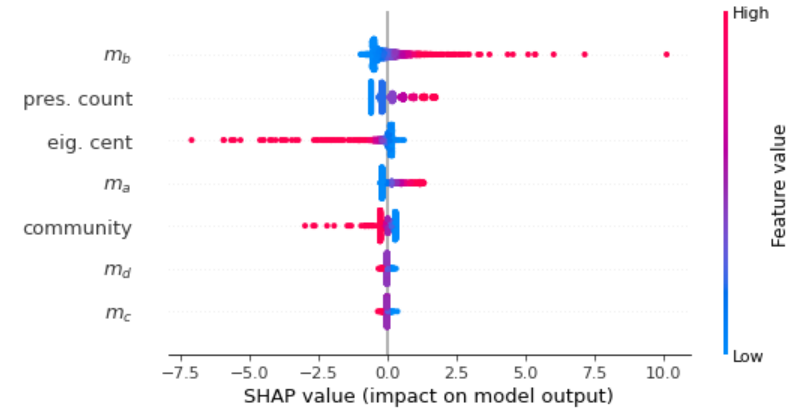}
    \caption{Equity-1}
    \label{fig:shap_1}
\end{subfigure}\\
\begin{subfigure}{\textwidth}    
    \centering
    \includegraphics[width=0.7\textwidth]{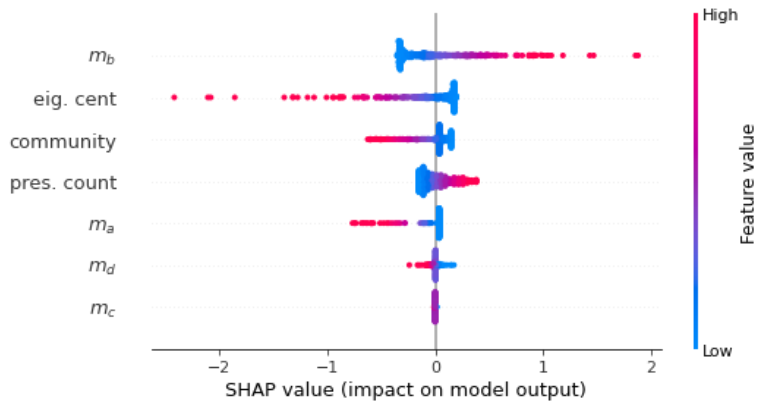}
    \caption{Equity-2}
    \label{fig:shap_2}
\end{subfigure}\\
\begin{subfigure}{\textwidth}
    \centering
    \includegraphics[width=0.7\textwidth]{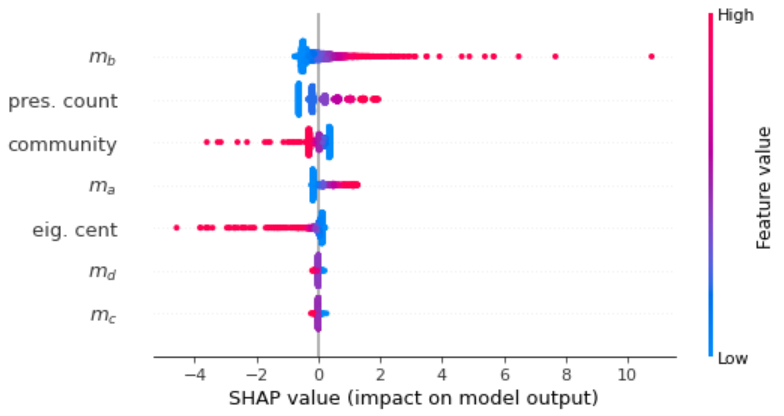}
    \caption{Equity-3}
    \label{fig:shap_3}
\end{subfigure}

\caption{SHAP feature importance for the different node importance measures.}
\label{fig:shap}
\end{figure*}

\begin{figure*}
    \centering
    \includegraphics[width=\textwidth]{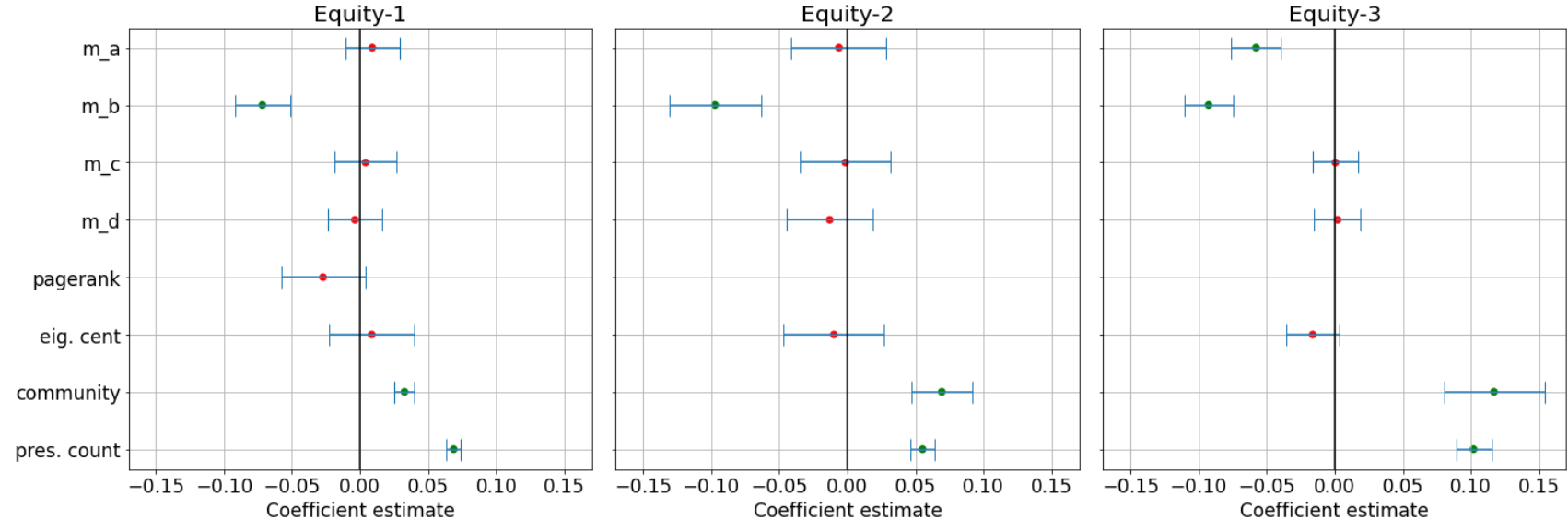}
    \caption{Classification model coefficients, with 95\% intervals indicated by the error bars. If a p-value is less than 0.05, it is coloured green, otherwise red.}
    \label{fig:class_coefs}
\end{figure*}

The precision and recall of the classifiers when applied to test sets are shown in table \ref{tab:pred_res}, alongside the performance of the null model. For all three datasets, the prediction showed reasonable precision and recall, which in all cases showed no overlap in the 95\% confidence intervals with the null model. Figure \ref{fig:feature_imps} shows the permutation importance for the different features. 
We see that the measure $m_b$ is by far the most important feature in the prediction across all three datasets considered. Although for equity-3 $m_a$ and eigenvector centrality are moderately important, as would be expected in a network with a single connected component, along with the presence count, none of the other node level measures are consistently important across all 3 datasets. This is consistent with our observations in figure \ref{fig:shap}, in which each dot represents a single SHAP explanation of the log-odds for a single observation by the feature the row of the plot corresponds to.
The features are ordered by the mean absolute value of the SHAP value for each feature. We see here that $m_b$ is also the most important feature on average for all three datasets, and that nodes with a high $m_b$ have a lower chance of remaining unchanged. In figure \ref{fig:class_coefs}, which shows the coefficients of the model, we see that $m_b$, community and presence count are the only features that show a consistent sign and approximate size of the parameter and also p-values of <0.001. 
We also see that $m_b$ has the largest magnitude\footnote{The features are standardised prior to use in the model, which allows for size comparison of the coefficients.} for both equity-1 and equity-2, which is in agreement with the feature importances in figure \ref{fig:feature_imps}. If the parameter $m_b$ is used as the only feature in the model, we see that the coefficient is consistently negative and significant, which indicates that nodes that are more important are less likely to subsequently transact. 


\begin{table}
    \centering
    \begin{tabular}{|c|c|c|c|}
    \hline
       Measure  & Equity-1 & Equity-2 & Equity-3  \\
       \hline
       Precision & 0.83  & 0.79  & 0.73  \\
       CI & (0.78,0.87) & (0.72,0.85) & (0.70,0.76) \\
       P (N.M.) & 0.54  & 0.60  & 0.57  \\
       CI & (0.47,0.59) & (0.52,0.69) & (0.53,0.62)\\
       Recall & 0.66  & 0.81  & 0.72 \\
       CI & (0.61,0.71) & (0.75,0.88) & (0.69,0.75) \\
       R (N.M.) & 0.49  & 0.51  & 0.50 \\
       CI & (0.29,0.60) & (0.21, 0.75) & (0.28,0.69)\\
       AUC & 0.66 & 0.61 & 0.76 \\
       CI & (0.64, 0.67) & (0.61, 0.63) & (0.75, 0.76) \\
       \small{AUC (N.M.)} & 0.5 &  0.5 & 0.5 \\
       CI & (0.47, 0.54) & (0.44, 0.55) & (0.47, 0.54) \\
       \hline
    \end{tabular}
    \vspace{0.1cm}
    \caption{Precision, recall and ROC AUC for the classification model for the 3 different datasets, presented alongside the same average precision and recall for the null model (N.M.) trials. The brackets denote the 95\% Confidence Intervals (CI). }
    \label{tab:pred_res}
\end{table}

When considering the exercise of predicting whether or not nodes are subsequently present from the different measures, in order to validate that the measure $m_b$ is the most predictive, we considered re-running the model with the measure $m_b$ removed. The results for this are found in table \ref{tab:pred_res_no_mb}. We see that although there is a drop in all precision scores, this only brings the model performance within the Confidence Interval range of the null model for the equity-2 dataset. The recall and ROC AUC also drop for all datasets, falling for both the equity-1 and equity-2 datasets to be within the Confidence Interval range of the null model. As expected, the model still retains some performance since some of the information captured by the measure $m_b$ is also captured by the other features, as shown by the correlations between the features. In this case, for equity-1 and equity-2, degree was found to be the most important of the remaining features and for equity-3, $m_a$ was the most important. The drop in model performance was least prominent for equity-3, which is as expected due to $m_b$ providing additional information in networks with disconnected components, which is not the case for equity-3.

\begin{table}
    \centering
    \begin{tabular}{|c|c|c|c|}
    \hline
       Measure  & Equity-1 & Equity-2 & Equity-3  \\
       \hline
       Precision  & 0.77 & 0.76 & 0.69  \\
       CI & (0.73,0.83) & (0.70, 0.83) & (0.66,0.73) \\
       P (N.M.)  & 0.67 & 0.70 & 0.57  \\
       CI & (0.62,0.71) & (0.63, 0.83) & (0.53,0.59)\\
       Recall & 0.40 & 0.5 & 0.69 \\
       CI & (0.36,0.44) & (0.43, 0.57) & (0.66,0.72) \\
       R (N.M.) & 0.52 & 0.51 & 0.49  \\
       CI & (0.27, 0.75) & (0.31, 0.74) & (0.38,0.64)\\
       AUC & 0.53 & 0.60 & 0.71 \\
       CI & (0.50, 0.57) & (0.53, 0.63) & (0.71, 0.72) \\
       AUC (N.M.) & 0.50 & 0.51 & 0.50 \\
       CI & (0.47, 0.54) & (0.45, 0.57) & (0.47, 0.52) \\
       
       \hline
    \end{tabular}
    \vspace{0.1cm}
    \caption{Precision and recall for the logistic regression for the 3 different datasets, presented alongside the same average precision and recall for the null model (N.M.) trials, when considering the prediction without the feature $m_b$. The brackets denote the 95\% Confidence Intervals (CI). }
    \label{tab:pred_res_no_mb}
\end{table}

For nodes which show a persistence in transacting between snapshots, we also considered whether the sign of the change to nodes' strength is predictable from the different node level features. The results of this are shown in table \ref{tab:pred_res_sign}. For this, we observe precision and ROC AUC scores for all three datasets which are not within the confidence intervals of the null model. However, the recall is never outside of the confidence intervals of the null model, so we cannot conclude that the sign of a change is predictable from our chosen node level features. In all cases, the `presence count' feature was the most important feature in predicting sign change.

\begin{table}
    \centering
    \begin{tabular}{|c|c|c|c|}
    \hline
       Measure  & Equity-1 & Equity-2 & Equity-3  \\
       \hline
       Precision & 0.80  & 0.84  & 0.93 \\
       CI & (0.75,0.84) & (0.77,0.90) & (0.91,0.95) \\
       P (N.M.) & 0.68  & 0.69  & 0.75  \\
       CI & (0.62,0.73) & (0.60,0.76) & (0.71,0.80)\\
       Recall & 0.66  & 0.68  & 0.75 \\
       CI & (0.61,0.71) & (0.61,0.76) & (0.71,0.78) \\
       R (N.M.) & 0.49  & 0.51  & 0.49  \\
       CI & (0.31,0.67) & (0.38, 0.65) & (0.27,0.71)\\
       AUC & 0.76 & 0.74 &  0.85 \\
       CI & (0.74, 0.76) & (0.73, 0.76) & (0.84, 0.85) \\
       AUC (N.M.) & 0.50 & 0.49 & 0.50 \\
       CI & (0.45, 0.54) & (0.41, 0.56) & (0.47, 0.54)\\
       \hline
    \end{tabular}
    \vspace{0.1cm} 
    \caption{Precision and recall for the logistic regression predicting the sign of the change in strength for the 3 different datasets, presented alongside the same average precision and recall for the null model (N.M.) trials, when considering the predicting the sign of the subsequent change in strength to a node. The brackets denote the 95\% Confidence Intervals (CI).}
    \label{tab:pred_res_sign}
\end{table}

\begin{figure*}
    \centering
    \includegraphics[width=\textwidth]{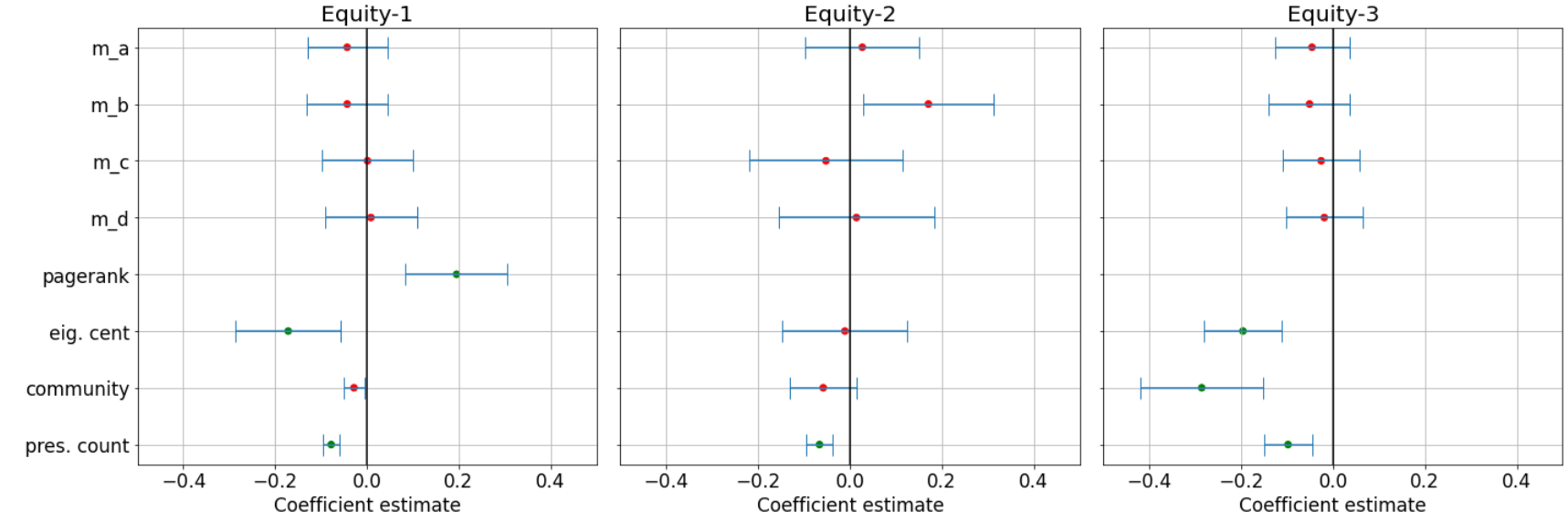}
    \caption{Regression model coefficients, with 95\% intervals indicated by the error bars. If a p-value is less than 0.05, it is coloured green, otherwise red.}
    \label{fig:reg_coefs}
\end{figure*}

Finally, we further considered whether the value of the change in strength is predictable from the node level features, by considering a regression of the features onto the value of the relative change in strength. The results for this are found in table \ref{tab:pred_res_regression}. 
\begin{table}
    \centering
    \begin{tabular}{|c|c|c|c|}
    \hline
       Measure  & Equity-1 & Equity-2 & Equity-3  \\
       \hline
       $R^2$ score & 0.102  & 0.111 & 0.103  \\
       CI & (0.08,0.16) & (0.08,0.19) & (0.082,0.152) \\
       $R^2$ (N.M.) & 0.118  & 0.112  & 0.107  \\
       CI & (0.102,0.132) & (0.097,0.130) & (0.100,0.117)\\
       \hline
    \end{tabular}
    \vspace{0.1cm}
    \caption{Coefficient of determination $R^2$ for a linear regression with endogenous value of the relative change in node strength, exogenous variables the node level features as used in the classification exercise. This is compared to a null model in which the relative change in node strength is randomly shuffled in 100 trials. We report the average and 95\% Confidence Intervals (CI).}
    \label{tab:pred_res_regression}
\end{table}
We see that for all three datasets, the Confidence Intervals for the $R^2$ score of the regression overlap with those for a null model in which node strength is randomly shuffled and that the $R^2$ values are higher for the null model. This means that we can conclude that the change in strength is not predictable in these networks. When looking at the coefficients of the regression model in figure \ref{fig:reg_coefs}, only eigenvector centrality and community show consistent sign of the coefficient across all three datasets, only presence count shows a high significance for the coefficients across the three datasets and in general many of the coefficient values are close to 0.


\subsection{Application to open source data: BaTIS dataset}
\label{sec:wto}
\begin{figure*}
    \centering
    \includegraphics[width=\textwidth]{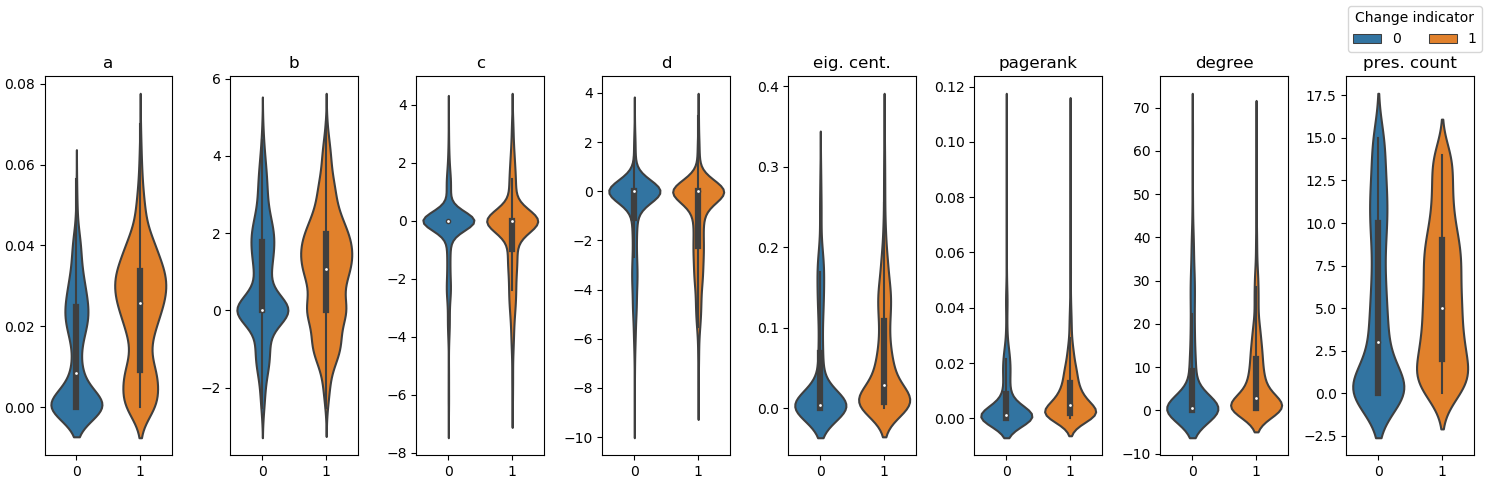}
    \caption{Distributions of the different node importance measures across edges which do subsequently change in comparison to those that don't, for the BaTIS trade dataset.}
    \label{fig:wto_boxplots}
\end{figure*}

\begin{figure}
    \centering
    \includegraphics[width=\linewidth]{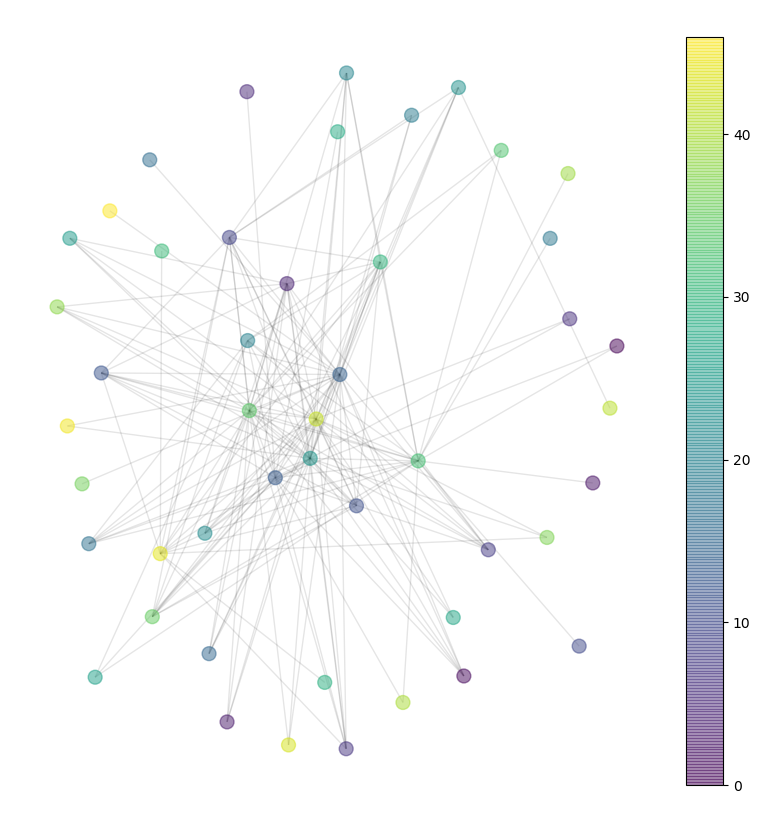}
    \caption{Network showing the initial snapshot for the BaTIS dataset, colours and numbers representing the ranking of the eigenvalue that localises to a given node.}
    \label{fig:wto_network}
\end{figure}

Since the equity transaction datasets explored above are not publicly available, here we present the results of applying our methods to the Financial Services segment of the BaTIS dataset. In contrast to the equity datasets, the BaTIS dataset has a natural persistence of activity (generally, countries that trade with each other continue to do so year on year) so instead of looking to predict whether or not a node will be present in the subsequent snapshot, we instead look to predict whether or not a node will show a change in strength in the subsequent snapshot\footnote{Specifically, we define a significant change to a node's strength as a change of more than 5\% between snapshots.}. Figure \ref{fig:wto_boxplots} shows the distributions of the different node level metrics for nodes that do subsequently change and those that don't \footnote{Note that a log transform has been applied to $m_b$, $m_c$ and $m_d$ due to these features spanning a few orders of magnitude}. 
We also observe how selection of the maximum eigencomponent for each node manifests itself in this dataset in figure \ref{fig:wto_network}, which shows the initial snapshot of the network with nodes coloured by the rank of the most relevant eigencomponent for that node. 

\begin{figure}
    \centering
    \includegraphics[width=\linewidth]{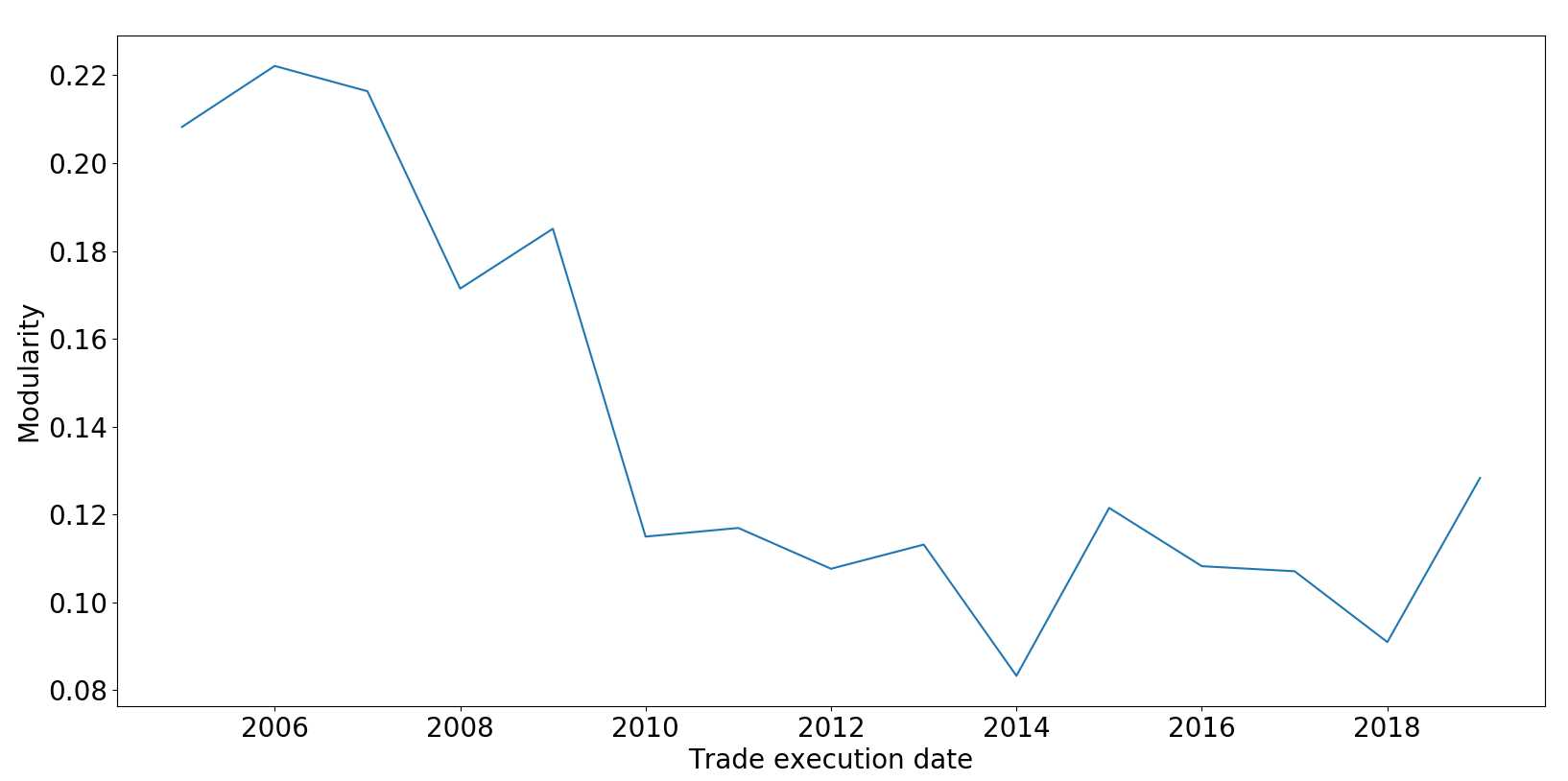}
    \caption{Modularity across time for the BaTIS dataset}
    \label{fig:wto_modularity}
\end{figure}

In figure \ref{fig:wto_modularity}
 we present the modularity across time for this dataset and we see that in comparison to the equity transaction datasets, the BaTIS dataset shows a much lower average modularity, and unlike the equity datasets which showed a stable modularity across time, we see a decreasing trend for this dataset. 
This is in agreement with what we see in figure \ref{fig:wto_network}, in which we see that the network has just one community, with a small number of peripheral nodes. These peripheral nodes tend to show lower eigenvalue rankings in comparison to the more densely connected core. Looking at figure \ref{fig:wto_boxplots}, we see that the measures $m_a$, $m_b$, eigenvector centrality and presence count show clear differences in the distributions for changing vs. unchanging nodes. The p-values for a two-sided t-test for the differences in the mean values for change vs. no change for each of the different measures are shown in table \ref{tab:p-values_wto}. Here we see that $m_a$ shows the most significant difference in the means, but $m_c$ and degree have p-values >0.05, suggesting that any difference we can visually observe for these variables is not unlikely to have occurred by chance. 

\begin{table}
    \centering
    \begin{tabular}{|c|c|}
    \hline
        Measure & BaTIS dataset  \\
        \hline
        $m_a$ & $ 2.62\times 10^{-47}$  \\
        $m_b$ & $6.34\times 10^{-6}$  \\
        $m_c$ & $5.83\times 10^{-2}$ \\
        $m_d$ & $1.71\times 10^{-3}$   \\
        Degree & $2.19\times 10^{-1}$  \\
        Eig. cent. & $3.51\times 10^{-7}$   \\
        Pagerank  & $2.79\times 10^{-4}$ \\
        Pres. count  & $1.46\times 10^{-4}$ \\
        \hline
    \end{tabular}
    \caption{p-values for a two-sided t-test for the differences in the mean values for nodes which change and nodes which don't, for each of the different node level measures, for the BaTIS dataset}
    \label{tab:p-values_wto}
\end{table}

\begin{figure}
    \centering
    \includegraphics[width=\linewidth]{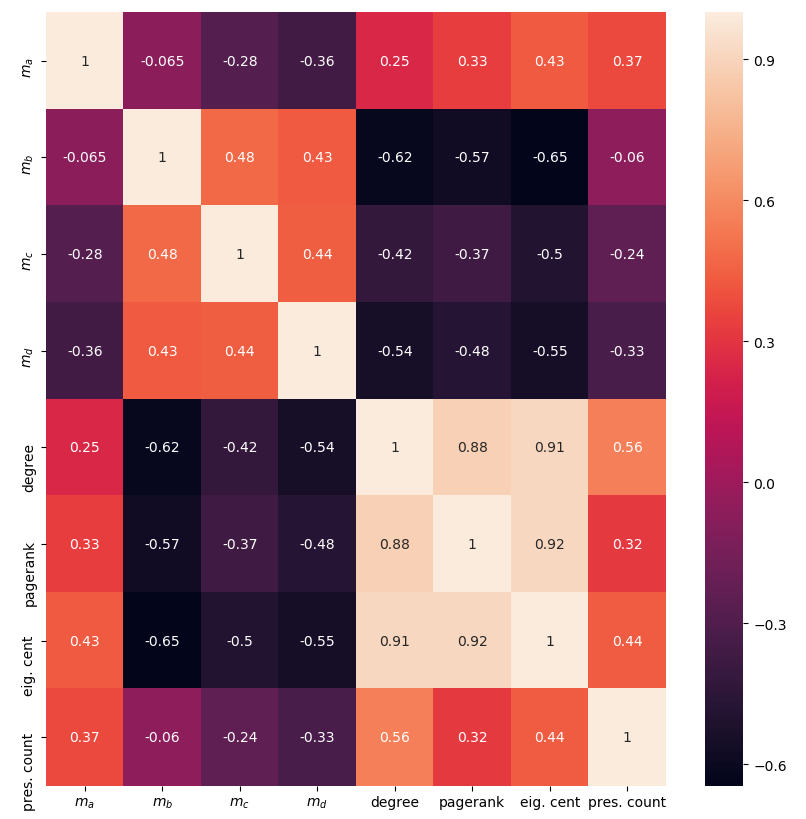}
    \caption{Pearson correlations for the BaTIS dataset.}
    \label{fig:wto_corrs}
\end{figure}

Now, prior to considering the role of node importance in change predictability for the BaTIS dataset, we first of all explore the correlations between the different measures used in the predictor, shown in figure \ref{fig:wto_corrs}. Here, we see large correlations between degree and the two centrality measures pagerank and eigenvector centrality, suggesting that these measures are only capturing the node degree as an indicator for importance. The measures $m_a$, $m_b$, $m_c$ and $m_d$ show no significant correlations with the other measures, although there is a reasonable negative correlation between $m_b$ and degree, eigenvector centrality and pagerank, which is as expected from our definition of structural importance, particularly in the case when there is only one community in the network. 

\subsubsection{Prediction experiments}
Following exclusion of correlated variables, the classifier shows good performance in table \ref{tab:wto_pred_res} on the test set, with a precision of 0.70, recall of 0.65 and a ROC AUC of 0.71. We see that although the precision and ROC AUC are better than the null model, the recall confidence intervals overlap with those of the dummy model so we cannot conclude that subsequent change to node strength is predictable from the node level features used for this dataset. 

\begin{figure}
    \centering
    \includegraphics[width=\linewidth]{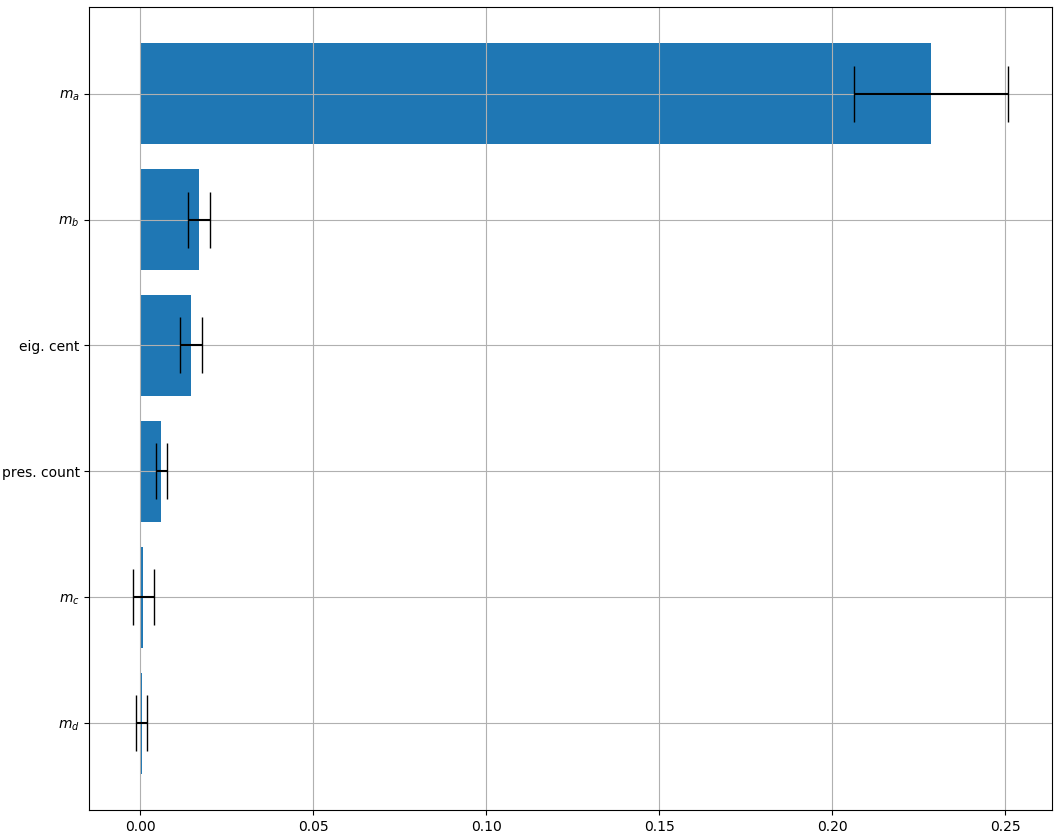}
    \caption{Logistic regression permutation importance for the different importance measures, for the BaTIS dataset.}
    \label{fig:wto_imps}
\end{figure}

\begin{figure}
    \centering
    \includegraphics[width=\linewidth]{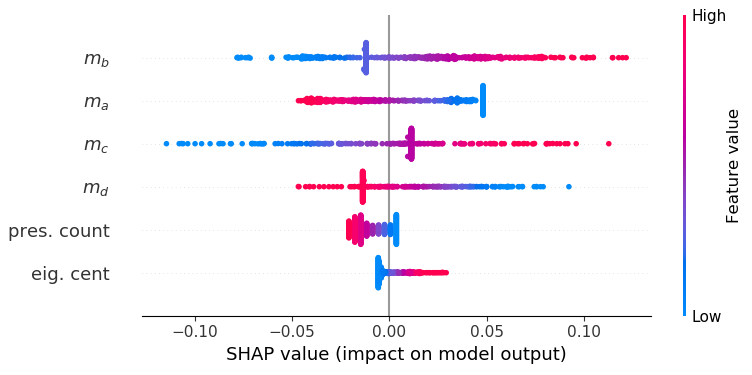}
    \caption{SHAP feature importance for the different node measures, for the BaTIS dataset.}
    \label{fig:wto_shap}
\end{figure}

\begin{figure}
    \centering
    \includegraphics[width=\linewidth]{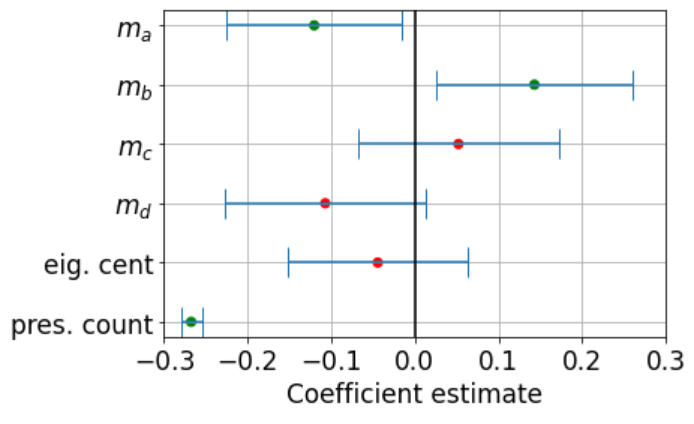}
    \caption{Classification model coefficients for the BaTIS dataset., with 95\% intervals indicated by the error bars. If a p-value is less than 0.05, it is coloured green, otherwise red.}
    \label{fig:wto_class_coeffs}
\end{figure}

Considering the feature importances in figure \ref{fig:wto_imps}, in contrast to the Equities transaction networks, we now see that the feature $m_a$ shows the largest importance, followed by $m_b$. This is as expected given the connected nature of the network, meaning that the measure $m_a$ is not impacted by disconnected components, as the leading eigenvalue will be the relevant eigenvalue for all nodes. We see similar results when looking at the SHAP values in figure \ref{fig:wto_shap} but we observe the most important feature to be $m_b$ closely followed by $m_a$. When looking at the coefficients in figure \ref{fig:wto_class_coeffs}, only eigenvector centrality does not show a significant coefficient and only $m_d$ and presence count show positive coefficients suggesting that higher values of these features would be indicative of nodes being more likely to change, whereas the other coefficients would show larger values for nodes that are less likely to subsequently change. 

\begin{table}
    \centering
    \begin{tabular}{|c|c|}
    \hline
       Measure  & BaTIS dataset  \\
       \hline
       Precision & 0.70 (0.63,0.76)  \\
       Precision (null model) & 0.52 (0.45,0.59) \\
       Recall & 0.65 (0.59,0.72)  \\
       Recall (null model) & 0.52 (0.36,0.66) \\
       ROC AUC & 0.71 (0.69, 0.71) \\
       AUC (null model) & 0.50 (0.46, 0.54) \\
       \hline
    \end{tabular}
    \caption{Precision and recall for the logistic regression for the BaTIS dataset, presented alongside the same average precision and recall for the Null model trials. The brackets denote the 95\% Confidence Intervals.}
    \label{tab:wto_pred_res}
\end{table}

We now consider the predictability of a change in sign of the strength, as considered for the equities networks. For this dataset, we see similar results to the equities data as shown in table \ref{tab:wto_pred_res_sign}, with good performance of the model in terms of precision and ROC AUC. However, we observe poorer performance for the recall showing overlap with the confidence intervals of the null model, meaning that again we cannot conclude that sign change is predictable from our node level features. For the case of sign prediction, the presence count was found to be the most important feature, closely followed by $m_a$.

\begin{table}
    \centering
    \begin{tabular}{|c|c|}
    \hline
       Measure  & BaTIS dataset  \\
       \hline
       Precision & 0.80 (0.72,0.89)  \\
       Precision (null model) & 0.66 (0.58,0.74) \\
       Recall & 0.65 (0.57,0.74)  \\
       Recall (null model) & 0.51 (0.39,0.64) \\
       AUC & 0.69 (0.65, 0.73) \\
       AUC (null model) & 0.50 (0.47, 0.52) \\
       \hline
    \end{tabular}
    \caption{Precision and recall for the logistic regression for the BaTIS dataset, presented alongside the same average precision and recall for the Null model trials, when considering predicting a change in sign for node strength. The brackets denote the 95\% Confidence Intervals.}
    \label{tab:wto_pred_res_sign}
\end{table}

\begin{figure}
    \centering
    \includegraphics[width=\linewidth]{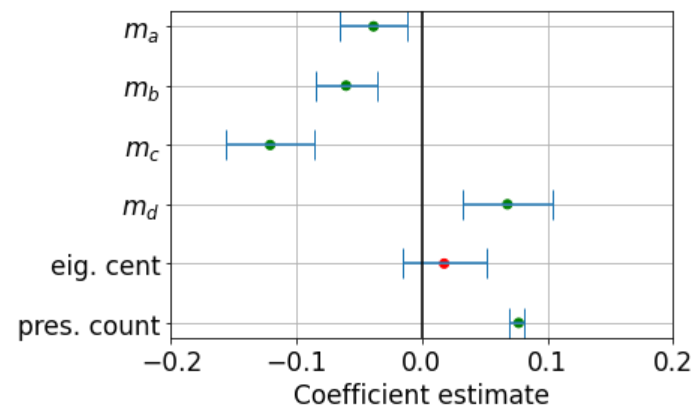}
    \caption{Regression model coefficient for the BaTIS dataset, with 95\% intervals indicated by the error bars. If a p-value is less than 0.05, it is coloured green, otherwise red.}
    \label{fig:wto_reg_coeffs}
\end{figure}

Finally, we consider the predictability of the value of the change in strength through the use of a linear regression model. The results for this are shown in table \ref{tab:wto_pred_res_regression}. In contrast to the equities dataset, the Coefficient of determination ($R^2$) suggest that the model has a reasonable prediction capability and although the confidence intervals for the dummy model are close, there is no overlap between these suggesting that this result is significant so we can conclude that the size of a change to a node's strength is predictable from our node level features for this dataset. The model coefficients for each of the features are shown in figure \ref{fig:wto_reg_coeffs}. Here we see that the significant coefficients are $m_a$, $m_b$ and presence count, and of these only $m_b$ has a positive coefficient suggesting that nodes with higher values of $m_b$ are more likely to show larger relative changes, whereas negative coefficients are seen for $m_a$ and presence count suggesting that nodes with lower values of these are more likely to show larger relative changes.
\begin{table}
    \centering
    \begin{tabular}{|c|c|c|c|}
    \hline
       Measure  & BaTIS dataset  \\
       \hline
       $R^2$ score & 0.69 \\
       CI & (0.66,0.72)  \\
       $R^2$ (N.M.) & 0.64  \\
       CI & (0.64,0.66) \\
       \hline
    \end{tabular}
    \vspace{0.1cm}
    \caption{Coefficient of determination $R^2$ for a linear regression with endogenous value of the relative change in node strength, exogenous variables the node level features as used in the classification exercise, for the BaTIS dataset. This is compared to a null model in which the relative change in node strength is randomly shuffled in 100 trials. We report the average and 95\% Confidence Intervals (CI)}
    \label{tab:wto_pred_res_regression}
\end{table}

\section{Discussion}
Through consideration of the full network spectrum, we present different ways of considering node importance in networks, with the aim of accounting for both community structure and hub nodes, which are key characteristics of many systems including financial networks. Motivated by several examples in the literature, to achieve this we demonstrate and make use of the most relevant eigenspectra entries in order to capture `community aware' node importance. The result of this is a measure which when applied to financial transaction networks, tells us how much a change to an individual node's strength, which in the context of equity transaction networks is their available funds or product, will impact the rest of the network. This sets our measure apart from centrality measures, as is it is defined in a temporal sense considering how the network will respond to changes.

By incorporating more than just the leading eigenpair, our measure is able to capture node importance in the context of complex structures, which makes our methods particularly suited for studying equity transaction networks. For these networks, which all show complex structures with both disconnected communities and `hub' nodes, we compare our measures of node importance to two commonly used centrality measures and also to degree, community label and the number of times a node has appeared historically, demonstrating that our measure is not simply acting as a proxy for these key node properties. When exploring whether static node importance is able to predict the presence of nodes in subsequent snapshots given features derived from the network history, we see that the measure $m_b$, which makes use of the eigenvector with the largest magnitude for each node, is the most important in determining the prediction for all three equities datasets.
Not only do our results demonstrate that $m_b$ is useful indicator of node importance in a static sense for networks with complex structures, they also provide evidence of the nodes in these equity networks having an evolution which depends on their importance when defined in this way. The latter of these observations is a useful insight for policy makers, as it motivates taking into account the full structure of these networks in determining which nodes to monitor more closely for their effects on the system. It also provides insights into the evolutionary properties of these networks which is interesting from a macro-economical perspective - we observe that more structurally important nodes are less likely to subsequently transact and given that these nodes tend to show positions in the network in which an impact to their strength would be spread across few counterparties, the observation of these nodes showing less frequent changes relates to the overarching stability of these networks.

Interestingly, if we compare our results on the equity datasets to application to a denser network of global trades in financial services (BaTIS), we observe poorer performance in the prediction of the presence or absence of changes and also the sign of any change. We also no longer see $m_b$ as the dominating feature in the prediction, suggesting that structural importance could be a unique property of sparse transaction networks. However, the value of a change to a node's strength for this dataset is predictable, with several node level features including $m_b$ being significant predictors. Further work is needed to understand the reasons for a lack of predictability of the value of a node's change in strength for the equity datasets, as one explanation for this could be the quality of the transaction reports or the methods of preprocessing applied to the data prior to our analysis.  

These observations contribute to the growing body of studies that provide insights into the evolution and stability of financial networks, for example the observations of Bardoscia et. al. \cite{Bardoscia1} that market practices that contribute to cyclical patterns tend to amplify distress. Further relevant to our work is Haldane et. al. \cite{Haldane2013}, in which it is noted that up until the 2008 crisis, the global financial system appeared to be self-regulating and self-repairing despite experiencing several exogenous shocks. However, in the crisis, enduring stress in the money markets was observed due to the interdependence of banks, who rationally sought to protect themselves from infection from other banks by hoarding liquidity. Our findings present us with a novel insight into the evolutionary behaviour of transaction networks for the capital markets, which we hope will motivate further research into the links between structural importance, network evolution and how these relate to market stability constraints.

Our experiments making use of logistic regression to predict subsequent transactions can be considered as a probabilistic model for the network dynamics. This complements approaches found in literature on econometric network models, where the much of the focus is on identification of models for macroeconomic time series given the wealth of price time series available for analysis. Approaches often make use of Vector Autoregressive models, which relate current observations of a variable with past observations of itself and past observations of other variables in the system \cite{lutkepohl2013vector}. Studies such as \cite{AHELEGBEY_16} have demonstrated how these models have good predictive accuracy and also offer a good representation of linkages between economic sectors, making them a useful tool for assessing systemic risk. An interesting further area of development for our methods would be to consider autoregressive approaches as an alternative model to the one we make use of in \ref{prob_model}.

Again in relation to econometric literature, additional insights could be gained by looking at the long-range effect of the network of transactions on stock prices and other financial variables, in a similar vein to the methods presented in \cite{Adamic_17}. In this paper the authors explore networks of traders of the S\&P 500 Stock index futures contracts and show how network variables preempt financial variables such as volume, duration and market liquidity measures, demonstrating the potential for trading networks in assessing liquidity supply and price formation influencing trading strategies. In relation to this, an interesting avenue for further work would be to compare the predictability observed to that obtained using correlation networks to analyse the more widely available data on stock prices, in a similar way to the comparison presented in \cite{Giudici_2016}, who demonstrate that a correlation based approach in combination with methods to analyse direct exposures provides a useful tool for assessing systemic risk. In their scenario of a limited dataset of direct exposures, the predictive power of correlation based approaches is significantly better than the approach making use of direct exposures.

It is worth noting that our measure of structural node importance, $m_b$, is not suitable for use on random networks, as in this case, no single entry of the eigenvector would be relevant for each node, so there is no guarantee that the eigenvector with the largest entry for each node is the correct part of the spectrum for that node. This restricts our method to the application of networks which are known to have a non-random structure. Further to this, we also note that the equity transaction datesets are sparse and may contain outlier values due to reporting errors. A useful further exploration would be to develop an equivalent to use of filtering methods from random matrix theory, which are widely used in identifying the relevant structure in correlation networks \cite{tumminello_2009,guo_2018,Barfuss_2016,ASTE2022655,DALY20084248,tumminello_05,massara_16}. Additional work could also include considering whether the nodes that are changing in these networks do so persistently, as this would allow us to gain deeper insights into the evolutionary behaviour of these networks and also an exploration of the resultant structural changes that occur when an important node changes. Our methods would also benefit from application to a large number of networks, both to further verify our observations and would be a useful tool for classifying networks according to their evolutionary properties.  

\section*{Availability of data and materials}
The datasets referred to as Equity-1, Equity-2 and Equity-3 in this paper, were extracted from a dataset of transaction reports collected by the FCA under MIFID II regulations. The datasets were used under agreement from the data owners at the Financial Conduct Authority for the current study and are not publicly available, so we also present comparable analysis for all investigations to an open source dataset. The dataset considered is the Financial Services segment of the Balanced Trade in Services (BaTIS) dataset, a complete and consistent trade in services matrix created by the OECD and WTO. Full details of the compilation methodologies can be found at \cite{wto_data}.

An implementation of the methods referenced in this paper can be found at \cite{my_code}. 
\section*{Funding}
This research did not receive any specific grant from funding agencies in the public, commercial, or not-for-profit sectors.
\bibliographystyle{unsrt}
\bibliography{bibliography.bib}
\end{document}